\documentclass[]{aastex61}

\newcommand{\MS}{\ifmmode{\,}\else\thinspace\fi{\rm M}\ifmmode_{\odot}\else$_{\odot}$\fi}
\newcommand{\LS}{\ifmmode{\,}\else\thinspace\fi{\rm L}\ifmmode_{\odot}\else$_{\odot}$\fi}
\newcommand{\RS}{\ifmmode{\,}\else\thinspace\fi{\rm R}\ifmmode_{\odot}\else$_{\odot}$\fi}

\shorttitle{High-precision Cepheid astrophysics}
\shortauthors{Pilecki et al.}

\begin{document}

\title{The Araucaria Project: High-precision Cepheid astrophysics from the analysis of variables in double-lined eclipsing binaries\footnote{This paper includes data gathered with the 6.5m Magellan Clay Telescope at Las Campanas Observatory, Chile.}
}

\correspondingauthor{Bogumi{\l} Pilecki}
\email{pilecki@camk.edu.pl}

\author[0000-0003-3861-8124]{Bogumi{\l} Pilecki}
\affiliation{Centrum Astronomiczne im. Miko{\l}aja Kopernika, PAN, Bartycka 18, 00-716 Warsaw, Poland}
\affiliation{Universidad de Concepci{\'o}n, Departamento de Astronom{\'i}a, Casilla 160-C, Concepci{\'o}n, Chile}

\author{Wolfgang Gieren}
\affiliation{Universidad de Concepci{\'o}n, Departamento de Astronom{\'i}a, Casilla 160-C, Concepci{\'o}n, Chile}
\affiliation{Millenium Institute of Astrophysics, Santiago, Chile}

\author{Grzegorz Pietrzy{\'n}ski}
\affiliation{Centrum Astronomiczne im. Miko{\l}aja Kopernika, PAN, Bartycka 18, 00-716 Warsaw, Poland}

\author{Ian B. Thompson}
\affiliation{Carnegie Observatories, 813 Santa Barbara Street, Pasadena, CA 91101-1292, USA}

\author[0000-0001-7217-4884]{Rados{\l}aw Smolec}
\affiliation{Centrum Astronomiczne im. Miko{\l}aja Kopernika, PAN, Bartycka 18, 00-716 Warsaw, Poland}

\author{Dariusz Graczyk}
\affiliation{Centrum Astronomiczne im. Miko{\l}aja Kopernika, PAN, Rabia{\'n}ska 8, 87-100 Toru{\'n}, Poland}

\author{M{\'o}nica Taormina}
\affiliation{Centrum Astronomiczne im. Miko{\l}aja Kopernika, PAN, Bartycka 18, 00-716 Warsaw, Poland}

\author{Andrzej Udalski}
\affiliation{Obserwatorium Astronomiczne Uniwersytetu Warszawskiego, Al. Ujazdowskie 4, 00-478 Warsaw, Poland}

\author{Jesper Storm}
\affiliation{Leibniz-Institut für Astrophysik Potsdam, An der Sternwarte 16, D-14482, Potsdam, Germany}

\author{Nicolas Nardetto}
\affiliation{Laboratoire Lagrange, UMR7293, Universit{\'e} de Nice Sophia-Antipolis, CNRS, Observatoire de la C{\^o}te d’Azur, Nice, France}

\author{Alexandre Gallenne}
\affiliation{European Southern Observatory, Alonso de C{\'o}rdova 3107, Casilla 19001, Santiago, Chile}

\author[0000-0003-0626-1749]{Pierre Kervella}
\affiliation{LESIA (UMR 8109), Observatoire de Paris, PSL, CNRS, UPMC, Univ. Paris-Diderot, 5 place Jules Janssen, 92195 Meudon, France}

\author{Igor Soszy{\'n}ski}
\affiliation{Obserwatorium Astronomiczne Uniwersytetu Warszawskiego, Al. Ujazdowskie 4, 00-478 Warsaw, Poland}

\author{Marek G{\'o}rski}
\affiliation{Universidad de Concepci{\'o}n, Departamento de Astronom{\'i}a, Casilla 160-C, Concepci{\'o}n, Chile}
\affiliation{Millenium Institute of Astrophysics, Santiago, Chile}

\author{Piotr Wielg{\'o}rski}
\affiliation{Centrum Astronomiczne im. Miko{\l}aja Kopernika, PAN, Bartycka 18, 00-716 Warsaw, Poland}

\author{Ksenia Suchomska}
\affiliation{Obserwatorium Astronomiczne Uniwersytetu Warszawskiego, Al. Ujazdowskie 4, 00-478 Warsaw, Poland}

\author{Paulina Karczmarek}
\affiliation{Obserwatorium Astronomiczne Uniwersytetu Warszawskiego, Al. Ujazdowskie 4, 00-478 Warsaw, Poland}

\author{Bart{\l}omiej Zgirski}
\affiliation{Centrum Astronomiczne im. Miko{\l}aja Kopernika, PAN, Bartycka 18, 00-716 Warsaw, Poland}

\begin{abstract}

Based on new observations and improved modeling techniques, we have reanalyzed seven Cepheids in the Large Magellanic Cloud. Improved physical parameters have been determined for the exotic system OGLE LMC-CEP-1718 composed of two first-overtone Cepheids and a completely new model was obtained for the OGLE LMC-CEP-1812 classical Cepheid. This is now the shortest period Cepheid for which the projection factor is measured. The typical accuracy of our dynamical masses and radii determinations is 1\%.
The radii of the six classical Cepheids follow period--radius relations in the literature. Our very accurate physical parameter measurements allow us to calculate a purely empirical, tight period--mass--radius relation that agrees well with theoretical relations derived from non-canonical models. This empirical relation is a powerful tool to calculate accurate masses for single Cepheids for which precise radii can be obtained from Baade--Wesselink-type analyses. The mass of the type-II Cepheid $\kappa$ Pav,  $0.56\pm{}0.08M_\odot$, determined using this relation is in a very good agreement with theoretical predictions.
We find large differences between the p-factor values derived for the Cepheids in our sample. Evidence is presented that a simple period--p-factor relation shows an intrinsic dispersion, hinting at the relevance of other parameters, such as the masses, radii, and radial velocity variation amplitudes. We also find evidence that the systematic blueshift exhibited by Cepheids, is primarily correlated with their gravity.
The companion star of the Cepheid in the OGLE LMC-CEP-4506 system has a very similar temperature and luminosity, and is clearly located inside the Cepheid instability strip, yet it is not pulsating.

\end{abstract}


\section{Introduction} \label{sec:intro}

Cepheids are probably the most important class of pulsating stars, having proved to be very useful in many different areas of astrophysics. Because of the period-luminosity (P-L) relation they obey, they are important distance indicators in the local universe, providing a fundamental step of the cosmic distance ladder and connecting our Milky Way galaxy to galaxies in the Local Group and beyond. They are also key objects for testing the predictions of stellar evolution and stellar pulsation theories. There are still many challenges to be faced, however.

Any metallicity dependence of the P-L relation (now known as the Leavitt law) will introduce systematic errors in the distance measurements and has to be accurately calibrated. Recent important progress on resolving this issue has been made by \citet{2018AA_Gieren_IRSB_SMC} but further independent tests are desirable.

Another way of using Cepheid variables as distance indicators is the Baade-Wesselink (BW) method, which uses the observed light, color, and radial velocity curves of Cepheids to determine their distances.
In any modern implementation of the BW method the dominant source of systematic uncertainty is the so-called projection factor (p-factor) which is needed to convert the observed radial velocities  into the pulsational velocities of the stellar surface. Only a handful direct measurements of this quantity have been reported in the literature to date, and the results are still quite confusing. Progress on knowledge of the p-factor, and its possible dependence on  stellar luminosity and  other parameters, is needed in order to use the BW method as a precision tool.

About 50 years ago after the work of \citet{1968QJRAS...9...13C} and \citet{1969MNRAS.144..485S}, it became apparent that the masses calculated from stellar evolutionary theory are higher than the values obtained from pulsation theory, typically by about 20\%. Various plausible solutions have been proposed to solve this discrepancy \citep{2006MmSAI..77..207B}, e.g., extension of the convective core during the main sequence evolution \citep{Cassisi_Salaris_2011ApJ.728L.43}, pulsation driven mass loss \citep{2011A&A...529L...9N}, and rotation \citep{2014A&A...564A.100A}. The determination of very accurate dynamical masses of Cepheids in eclipsing binary systems has been the key to the solution of this problem.

The so-called K-term was originally measured as a negative (blueshifted) residual Cepheid velocity in our Galaxy \citep{1974A&AS...15....1W,1994A&A...285..415P}. Since it is not possible that all the observed Cepheids are moving towards us (with a velocity of about 1-2 km/s), the best explanation is that the effect arises in the Cepheid atmospheres \citep{1995ApJ...446..250S,2009A&A...502..951N}. Although pulsations may contribute to this effect, it is probably mostly caused by  convective motions \citep{Vasilyev2017_2dcep_model_A&A.606.140}. A similar {\em convective blueshift} is observed for non-pulsating late-type stars as well.

Detached eclipsing double-lined spectroscopic binaries (SB2) provide us with the opportunity to directly and accurately measure stellar parameters such as mass, luminosity, and radius. We can also measure the distance to such systems using almost purely geometrical methods. 
Although a number of Cepheids in binary systems in our Galaxy were known, for many years no Cepheid in an eclipsing system was found. The breakthrough came with the microlensing surveys, MACHO \citep{macho2002alcock} and OGLE \citep{ogle2015udalski}. Some candidates in the Large Magellanic Cloud (LMC) were proposed, but a spectroscopic confirmation was needed in each case to prove the membership of the Cepheids in genuine binary systems.

OGLE LMC-CEP-0227 was the first Cepheid in an eclipsing binary system to be confirmed spectroscopically  \citep{cep227nature2010} and reanalyzed by \citet{cep227mnras2013} using a more advanced method and additional data.
Following OGLE LMC-CEP-0227, five other Cepheids have been confirmed and their parameters measured: OGLE LMC-CEP-1812 \citep{cep1812apj2011}, OGLE LMC-CEP-1718 A and B (so far the only known system composed of a pair of classical Cepheids  \citep{cep1718apj2014}), OGLE LMC-CEP-2532 \citep{cep2532apj2015}, and OGLE LMC562.05-9009 \citep{cep9009apj2015}. To this list a system containing a peculiar W~Virginis star component, OGLE LMC-T2CEP-098, has been added recently \citep{t2cep098apj2017}. A short summary of this subject was presented by \citet{2017EPJWC.15207007P}. For all these Cepheids in double-lined eclipsing binary systems, masses, radii, and other physical parameters have been measured with a precision of typically 1-3\% which has provided an enormous improvement over  previous determinations of these parameters for Cepheids in the Milky Way or the Magellanic Clouds.

Significant recent progress towards a physical understanding of the p-factor in Cepheids has been made. According to a recent theoretical study by \citet{2017A&A...597A..73N} the p-factor is composed of three parts: one is related to a geometric projection and the limb-darkening, while two others are related to the dynamical structure of the pulsating atmospheres and in particular to the atmospheric velocity gradient. There are also different observational approaches for the determination of p-factors, with differing advantages. A recent statistical analysis \citep{Storm_2011_pfac_AA.534.94}  has strengthened previous indications that the value of the p-factor shows a systematic trend with the pulsation period of the Cepheids, but newer results from direct methods \citep{Kervella2017_RS_Pup,2017A&A...608A..18G} give no clear answer on this dependence.

The largest number of direct measurements of p-factors now arises from interferometric studies and the SPIPS (SpectroPhoto-Interferometry of Pulsating Stars) modeling -- a generalization of the Baade-Wesselink method \citep{Merand_2015_spips_AA.584.80}. This methodology has been used to measure p-factors for five classical Cepheids \citep{Breitfelder2016_9_CCEP,Kervella2017_RS_Pup} and one type-II Cepheid \citep{Breitfelder2015_K_Pav} in our Galaxy. Unfortunately, in the SPIPS method there is a degeneracy between the p-factor and the distance to the star, which limits the precision of the p-factor determinations. This problem is expected to be resolved with the accurate parallaxes expected from the Gaia mission.
Measurements obtained from the analysis of Cepheids in eclipsing binary systems, although not as numerous at the present time, are the only ones which provide distance-independent geometrical p-factors.

This work is divided into two segments. The first one is focused on  individual targets, including improved models with better accuracy and precision, as well as the measurements of  new parameters (e.g. the p-factor). The second segment is dedicated to the analysis of the whole sample, including relations between the measured parameters, such as  the mass-radius-period relation for pulsating stars.

\section{Objects}

\begin{figure}
\begin{center}
  \resizebox{0.9\linewidth}{!}{\includegraphics{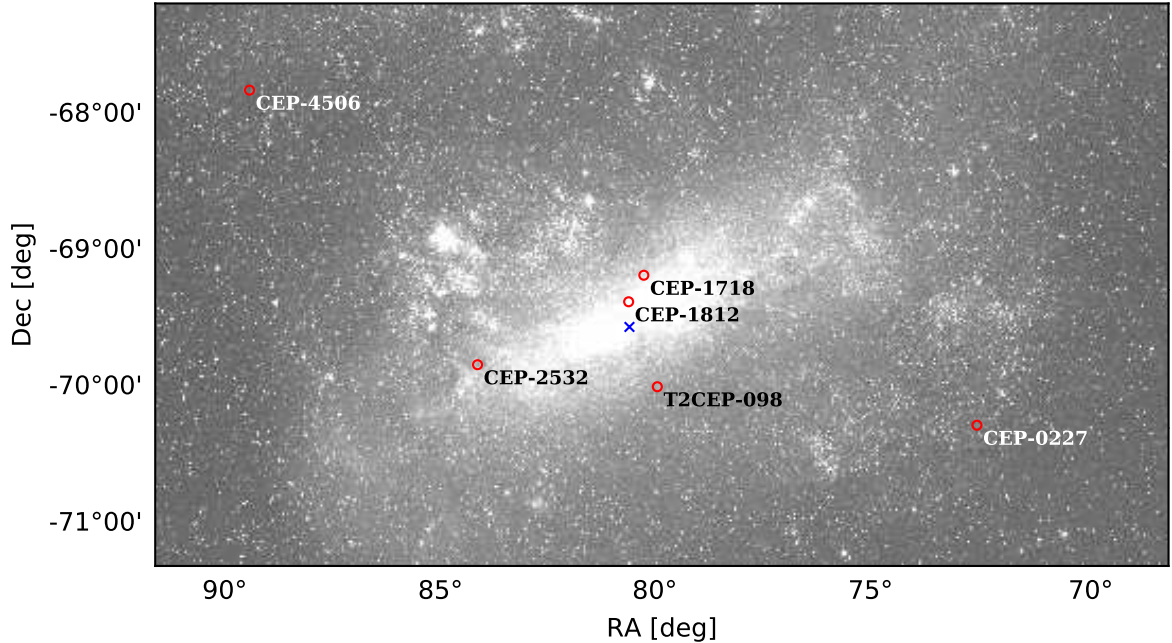}} \\
\caption{LMC view and the positions of the analyzed systems ({\em OGLE LMC} prefix was omitted in the names for clarity). CEP-4506 refers to the system previously known as LMC562.05.9009.  The galaxy center is marked with blue $\times$. Different text colors are used for better contrast only.
\label{fig:starsonsky}}
\end{center}
\end{figure}

In this study, we have checked and revised the parameters of six Cepheids in eclipsing binary systems in the LMC that have been previously published by our group. Their position in the sky within the LMC is presented in Fig.~\ref{fig:starsonsky}. 
Most of them are located close to the galaxy center (names in black), while two are found at a significantly higher distance (names in white).
A re-analysis of these Cepheids was warranted because we have been collecting many new observations over the past few years, while simultaneously improving our analysis tools (see section 3).

Among these six classical Cepheids (CC), three pulsate in the fundamental (F) mode and three in the first-overtone (FO) mode. There is also one fundamental mode pulsator (OGLE LMC-T2CEP-098), originally classified as a peculiar W Virginis star, which has characteristics more similar to Anomalous Cepheids than to classical or type-II Cepheids (T2C). Basic information about all of these objects is presented in Table~\ref{tab:basic}. 

One studied object (OGLE LMC562.05.9009) was later included in the OGLE Collection of Variable Stars (OCVS), with the name OGLE LMC-CEP-4506. Although it may spur some confusion, for consistency we will use the latter designation in this paper. As we use the OCVS designations for all of the objects, we will omit the {\em OGLE} prefix in the text, with the exception of the table captions.

\begin{deluxetable}{lccccccc}
\tablecaption{Basic data for the Cepheids in eclipsing binaries discussed in this paper \label{tab:basic}}
\tablewidth{0pt}
\tablehead{
\colhead{Object} & \colhead{mode} & \colhead{$P_{puls}$} & \colhead{$P_{orb}$} & \colhead{R.A.} & \colhead{Dec} & \colhead{$V_J$}  & \colhead{$I_C$}}
\startdata
\colhead{OGLE ID} &       & days        & days  & hh:mm:ss.ss   &$\pm$dd:mm:ss.s & mag   & mag    \\
\hline
LMC-CEP-1718$^a$  & FO+FO & 1.964+2.481 & 412.8 & 05:21:54.93   & -69:21:50.3   & 15.202 & 14.519 \\
LMC-CEP-1812      & F    & 1.313       & 551.8 & 05:23:07.70   & -69:33:49.9   & 16.693 & 16.054 \\ 
LMC-CEP-0227      & F    & 3.797       & 309.7  & 04:52:15.69   & -70:14:31.3   & 15.291 & 14.408 \\
LMC-CEP-2532      & FO    & 2.035       & 800.4 & 05:36:04.48   & -70:01:55.6   & 17.299 & 15.758 \\
LMC-CEP-4506$^b$  & F    & 2.988       & 1550  & 05:53:29.36   & -67:53:59.4   & 15.486 & 14.770 \\
LMC-T2CEP-098$^c$ & F    & 4.973       & 397.2 & 05:20:25.00   & -70:11:08.7   & 14.671 & 14.374 \\
\enddata
\tablecomments{Source references: OGLE-III \citep{2008AcA....58..163S}, OGLE-IV \citep{2015AcA....65..297S}. F stands for fundamental mode and FO for first-overtone.}
\tablenotetext{a}{System consists of two Cepheids; modes and periods for both are given.}
\tablenotetext{b}{Known before as OGLE LMC562.05.9009.}
\tablenotetext{c}{The ID refers to the original OGLE classification, but the star is an outlier, which has more in common with Anomalous Cepheids \citep{t2cep098apj2017}. }
\end{deluxetable}

All of the data used for the analysis of the objects presented in this paper are available  on a dedicated webpage\footnote{http://araucaria.camk.edu.pl/p/apcep}.

\section{Method and analysis} \label{sec:method}

For a detailed description of the method and the analysis steps, we refer the reader to \citet{cep227mnras2013} and the papers on other Cepheids that follow. Here we present  a short summary of the subject.

Radial velocities (RV) are measured using the Broadening Function method implemented in the {\tt RaveSpan} code written by B. Pilecki \citep[see Appendix in][]{t2cep098apj2017} with template spectra taken from the synthetic spectra library of \citet{2005A&A...443..735C}. Orbital solutions are then obtained using the same program. In {\tt RaveSpan} the pulsations are included and fitted, superimposed on the orbital motion. As a result, we have the masses and the system size (scaled by the inclination), the orbital parameters and the pulsational RV curve. The latter is used to model the radius change after scaling with the p-factor.

The photometric data were analyzed using a pulsation-enabled eclipsing modeling tool based on well-tested {\tt JKTEBOP} code \citep{jktebop2004southworth} modified to allow the inclusion of pulsation variability. We generate a two-dimensional light curve that consists of purely eclipsing light-curves for different pulsating phases. A one-dimensional light curve is then generated (interpolated from the grid) using a combination of pulsational and orbital phases calculated using the Cepheid and system ephemerides.

From the photometric solution, we have measurements of the period,  time of the primary minimum,  inclination,  fractional radii,  eccentricity,  argument of periastron, surface brightness ratios, third light (if present), and the value of the p-factor. Various limb darkening parameters are also tested.

Depending on the system, some parameters are taken exclusively from the spectroscopic or photometric solution, although they can be determined from both, e.g. period, time of the primary minimum, eccentricity and the argument of periastron. This is especially important where some information is missing in one data set, for example, when due to the system configuration only one eclipse per cycle can be observed. These issues will be indicated on an  object by object basis in the following sections.

To obtain optimal solutions and error estimates, standard Monte Carlo and Markov Chain Monte Carlo (MCMC) samplings are used.

\section{Updated models} \label{sec:update}

For most of the systems, we have been able to collect additional data after the latest publication describing that object. 
For LMC-CEP-1812, only a preliminary study was made and here we present the results from a completely new analysis. For LMC-T2CEP-098, we have not yet collected new data and present here only slightly revised physical parameters based on a new estimate of the reddening.

For the other systems, we have collected from a few to almost forty new spectra. In all cases, all the data were reanalyzed in a uniform way to ensure  consistency. There also have been  several improvements developed since the first publications which are now incorporated as detailed in the following sections. The new data and better data treatment have resulted in a significant improvement in the accuracy of the solutions for the systems. This is both because of the lower statistical error (more of better quality measurements) and because some sources of systematic errors have been  removed.

\subsection{OGLE LMC-CEP-1718} \label{subsec:cep1718}

This very interesting object, consisting of two FO Cepheids, was first detected by \citet{2008AcA....58..163S} in the OGLE data and later confirmed spectroscopically by \citet{cep1718apj2014}. A preliminary analysis was made in the latter, but the results suffered from various issues. The complicated nature of the system and the poorly covered pulsational and orbital radial velocity curves made it difficult to disentangle the velocities correctly (only 38 spectra were available, with 22 parameters to fit). In addition, only one eclipse is seen in the system.

\begin{figure}
\centering
\includegraphics[width=0.6\textwidth]{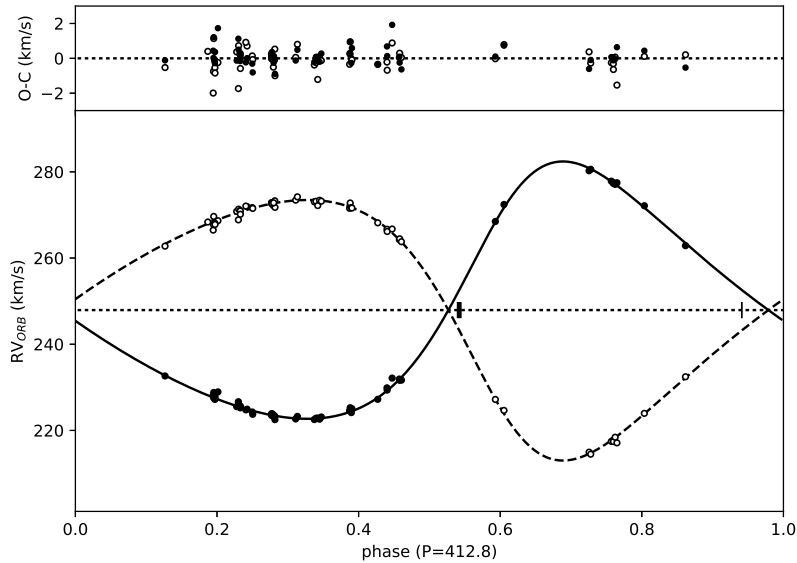}
\caption{Orbital radial velocity curve for LMC-CEP-1718 with the pulsational variabilities of both Cepheids removed. Short vertical lines mark the phases of conjunction. The eclipse (of the secondary) is seen in the light curve only at  inferior conjunction (thick line). Because of the system configuration (see Fig.~\ref{fig:1718_stars2}), there are no eclipses at the superior conjunction (thin line).
\label{fig:1718_orb}}
\end{figure}

\begin{figure}
\centering
\includegraphics[width=0.48\textwidth]{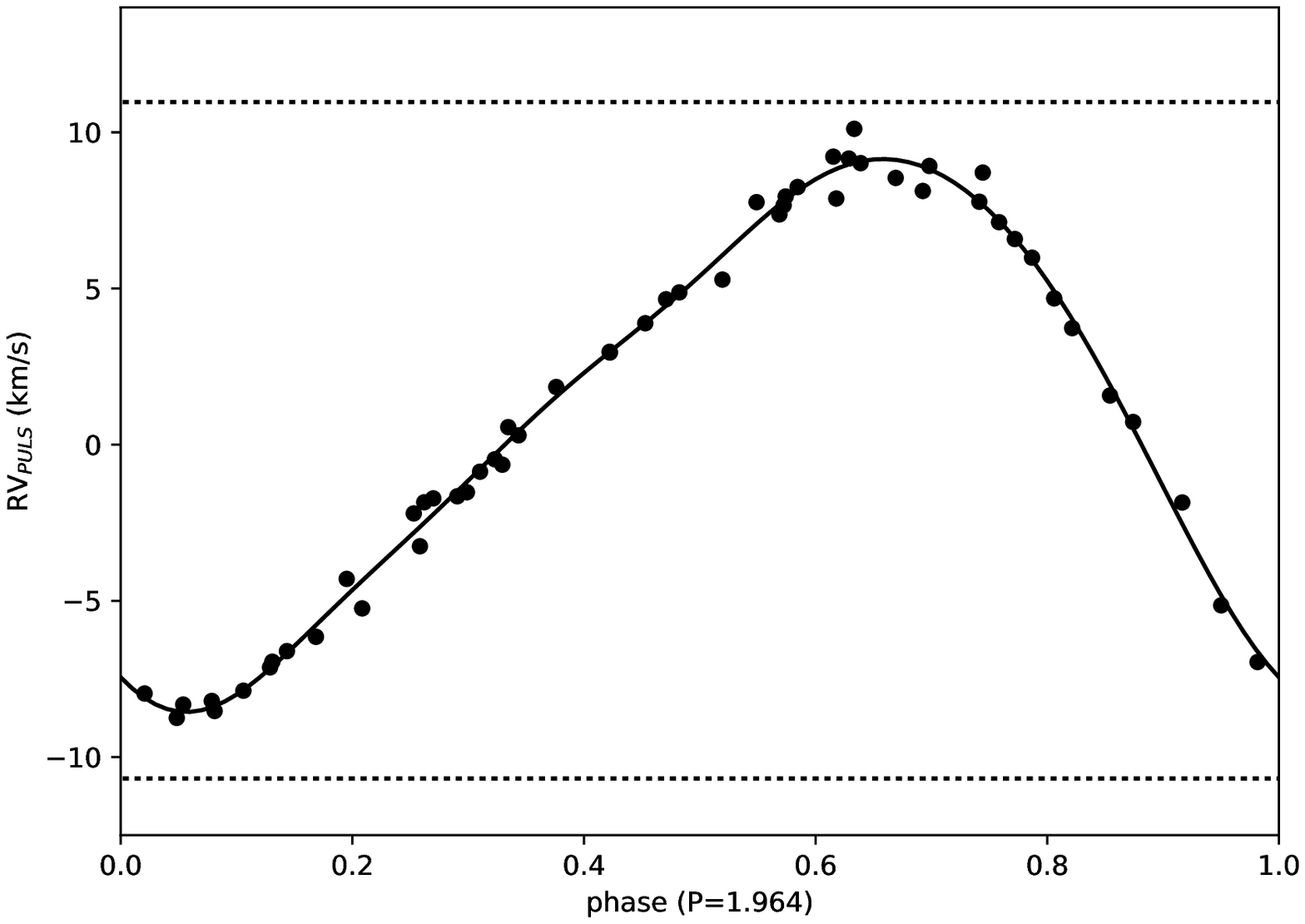}
\includegraphics[width=0.48\textwidth]{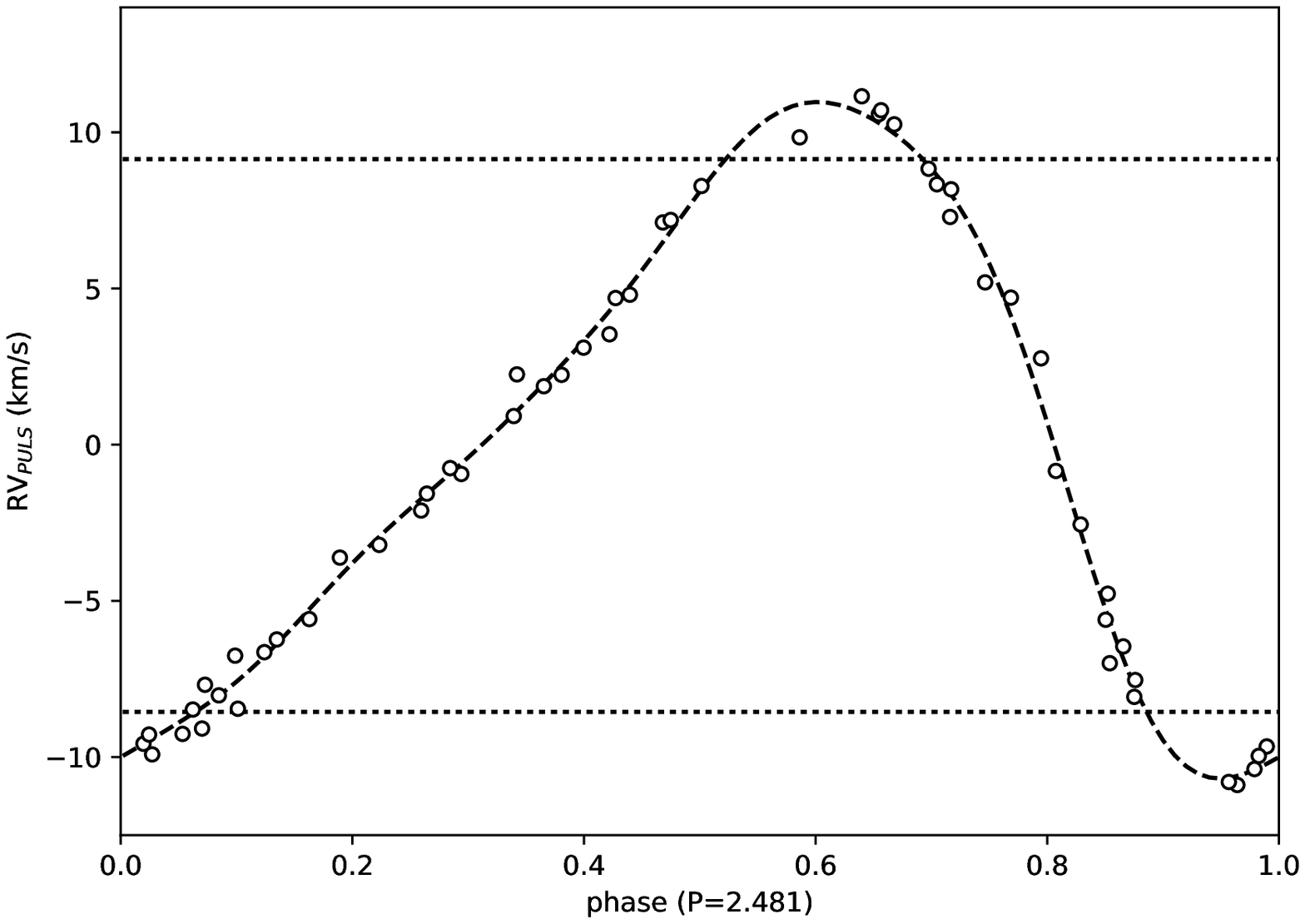}
\caption{Pulsational radial velocity curves of component A ({\em left}) and B ({\em right}) of the LMC-CEP-1718 system, plotted in the same vertical range. Dotted horizontal lines mark the amplitude of the other component for comparison.
\label{fig:1718_puls}}
\end{figure}

We have now collected a total of 58 spectra (22 with HARPS, 35 with MIKE, and 1 with UVES). The addition of new spectra allowed us to make a more detailed analysis of the radial velocity curves and correctly disentangle the pulsational and orbital motions of the stars -- see Fig.~\ref{fig:1718_orb} and \ref{fig:1718_puls}. 
Instrumental shifts were also taken into account in the analysis yielding lower scatter of the residual data -- the MIKE spectra had to be shifted by -900 m/s in regard to the HARPS spectra. The orbital solution is presented in Table~\ref{tab:1718_spec}.

Note that we do not differentiate here between the real instrumental shift and any other phenomenon that affects velocities in a similar way. For example, small differences in effective wavelength ranges (weighted by signal to noise ratio) between the instruments may also add to these shifts.


\begin{deluxetable}{lr@{ $\pm$ }lc}
\tablecaption{Orbital solution for OGLE LMC-CEP-1718  \label{tab:1718_spec} }
\tablewidth{0pt}
\tablehead{
\colhead{Parameter} & \multicolumn{2}{c}{Value} & \colhead{Unit}
}
\startdata
$\gamma_1$      &  248.52    &  0.11  &  km/s   \\ 
$\gamma_2$      &  247.75    &  0.15  &  km/s   \\
$T_0$ (HJD)     & \multicolumn{2}{c}{2456477.99} &  d      \\
$a \sin i$      &  472.53    &  1.1   &  $R_\odot$ \\ 
$m_1 \sin^3 i$  &    4.17    &  0.05  &  $M_\odot$ \\ 
$m_2 \sin^3 i$  &    4.12    &  0.04  &  $M_\odot$ \\ 
$q=m_2/m_1$     &    0.988   &  0.005 &  $M_\odot$  \\ 
$e$             &    0.269   &  0.004 &  -      \\
$\omega$        &  305.2     &  1.0   &  deg  \\
$K_1$           &   29.89    &  0.10  &  km/s \\ 
$K_2$           &   30.24    &  0.11  &  km/s \\ 
rms$_1$         & \multicolumn{2}{c}{0.55} &  km/s \\
rms$_2$         & \multicolumn{2}{c}{0.64} &  km/s \\
\enddata
 \tablecomments{ $T_0$ is calculated from the epoch of the secondary minimum: 2456701.7737.}
\end{deluxetable}

To obtain a consistent solution, both period and a reference time of orbital phase 0.0\footnote{We refer here to phase 0.0 in a generalized model that can be used consistently for systems with apsidal motion. It coincides with the time of the primary minimum for circular orbits and if $\omega=90^{\circ}$.} ($T_0$) were taken from the analysis of the light curve, as they can be calculated reliably, even without a good light curve model. $T_0$ was iteratively calculated from the epoch of the secondary minimum ($T_{sec}$ = 2456701.7737) and the current orbital solution (eccentricity $e$ and argument of periastron $\omega$) until  convergence was obtained. This differs from the previous analysis of the system, where $T_0$ and $T_{sec}$ were fitted separately, potentially leading to incoherent results.

The new solution is consistent with the old model with parameters not differing more than $5\%$, but with improvements in accuracy and precision.
The largest difference, when compared to the old model, is between the masses that suffered from an error in the calculations. The new and correct masses are higher by about $28\%$ (including $\sim 3\%$ due to the improved modeling and new data -- compare with the masses in \citealt{2017EPJWC.15207007P}), making them similar to the masses of the other Cepheids in binary systems presented below.
The semi-major axis is now larger by about $5\%$, which has increased the scale of the system and the Cepheids radii by the same amount.
The value of the determined mass ratio ($q=0.988 \pm 0.005$) is a strong indication that the stars are not of the same mass, with the primary being more massive by about 1\%. This observation is important to understand different properties of the Cepheid components and their periods in particular.

With the additional observations it was possible to fit separate systemic velocities for the two components ($\gamma_1$  and $\gamma_2$). The difference between them is a good estimate of the difference between the K-terms of the stars. Our results suggest that the lower mass, longer period Cepheid is  more blueshifted by $0.83 \pm 0.13$ km/s than the primary short period component. Such a residual K-term is interesting as it can be used to study the differences in the atmospheres of these otherwise  similar objects.

\begin{figure}
\begin{center}
  \resizebox{0.7\linewidth}{!}{\includegraphics{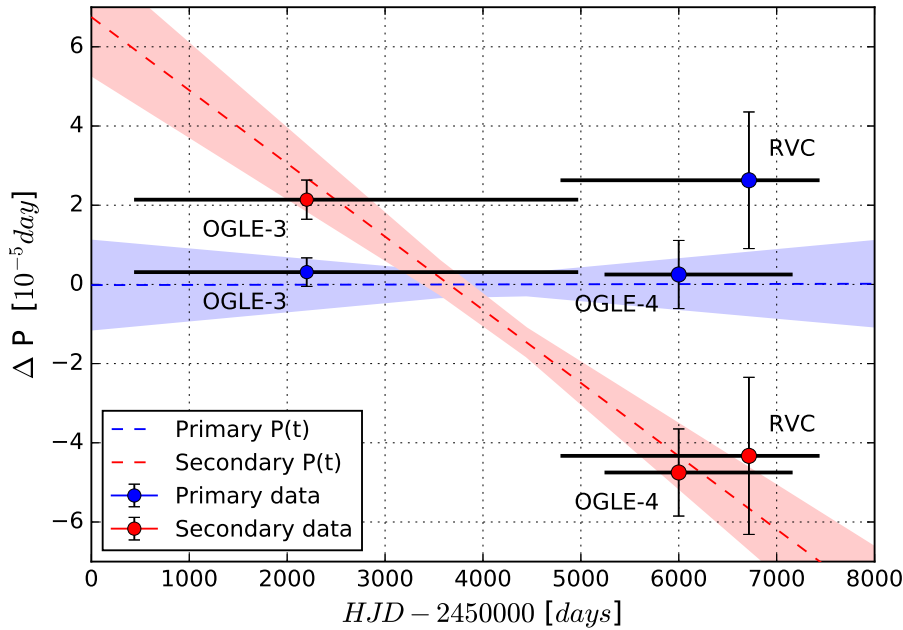}} \\
\caption{Period measurements for both Cepheids of the LMC-CEP-1718 system. $\Delta P$ is calculated with the reference to periods determined from OGLE-III and IV data together, i.e. $P = 1.9636627$ days and $P = 2.4809143$ days for the primary and secondary, respectively. A linear period change fitted to both datasets is shown as dashed lines, with the colored areas representing the uncertainties.
Horizontal solid lines show the timespan of the individual data sets, while the points mark their weighted averages and the measured periods (with errors). RVC marks periods obtained from the radial velocity curves.
\label{fig:1718_per}}
\end{center}
\end{figure}

For the first time, we have also checked for any period change in the component Cepheids. Using combined OGLE-III and OGLE-IV data for the secondary (longer period) component we have detected a significant period decrease at a rate of $\dot{P_2} = (-18 \pm 3) \times 10^{-9}$, whereas the primary shows no period change with  $\dot{P_1} = (0.0 \pm 2.3) \times 10^{-9}$. Using only OGLE-III data, we obtain similar results: $\dot{P_2} = (-12 \pm 5) \times 10^{-9}$ and $\dot{P_1} = (-5 \pm 5) \times 10^{-9}$, but with higher uncertainties.  Because of this detection of  period change in the secondary the pulsation periods were also fitted in the RV solution, while in general they have been taken from the light curve analysis. These RV periods are in general consistent with the measurements of  $\dot{P_1}$ and $\dot{P_2}$, but the $P_{RV}$ for the primary  is somewhat higher than expected perhaps indicating a very slow period increase. More data will be needed to confirm this possibility.
The period measurements are presented in Fig.~\ref{fig:1718_per}.
 
The modeling of the light curve of LMC-CEP-1718 is not only complicated because of the two pulsational variabilities superimposed on the eclipsing light curves, but also because only one and shallow eclipse (secondary) is visible in every cycle. This considerably limits the amount of information we can obtain from the modeling and makes it necessary to use all the other sources we have.
In particular, the eccentricity and the argument of periastron are well determined from the analysis of the radial velocity curves and we use their fixed values in the light curve modeling to lower the number of free parameters.
As a new orbital solution was obtained for this work, the light curve analysis was repeated as well.

In our previous study the surface brightness ratio of the stars was fitted, but as we do not see both eclipses the uncertainty in this parameter is very high. To avoid this problem we have decided to take advantage of both components being Cepheid variables. From the I-band period-luminosity relation \citep{2015AcA....65..297S} we can obtain the light ratio of the stars as $L_{21}^{IC} = 1.361 \pm 0.015$, and from the empirical period-radius relation \citep{2002AstL...28..589S} the radii ratio of $R_{21} = r_2/r_1 = 1.19 \pm 0.08$. The theoretical period-radius relations \citep{Bono2001ApJ...552L.141B} give the same value, but with much lower uncertainties. Then the I-band surface brightness ratio can be derived:

$$ j_{21}^{IC} = L_{21} / R_{21}^2 = 0.96 \pm 0.13 $$

\begin{figure}
\centering
\includegraphics[width=0.49\textwidth]{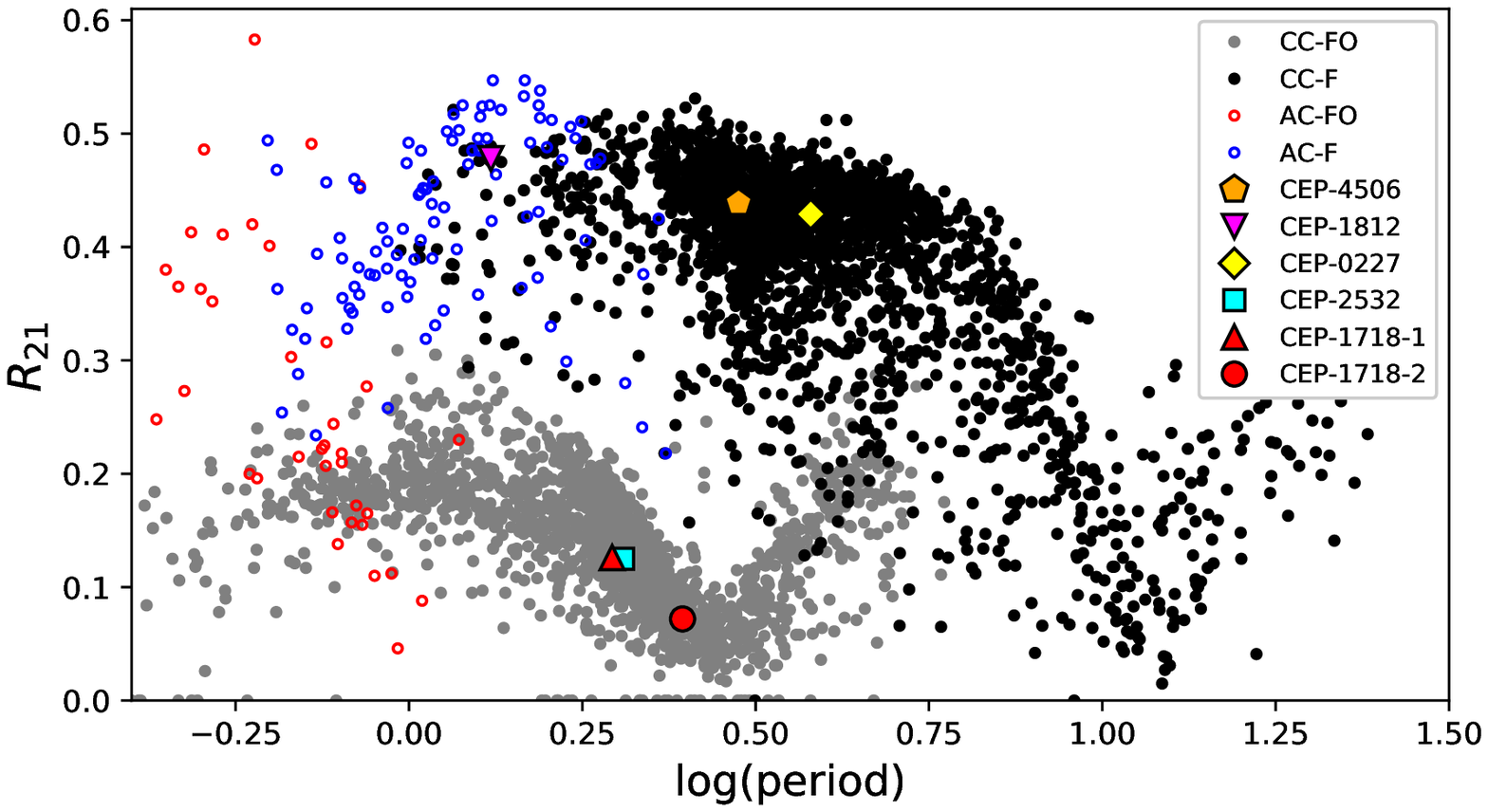}
\includegraphics[width=0.49\textwidth]{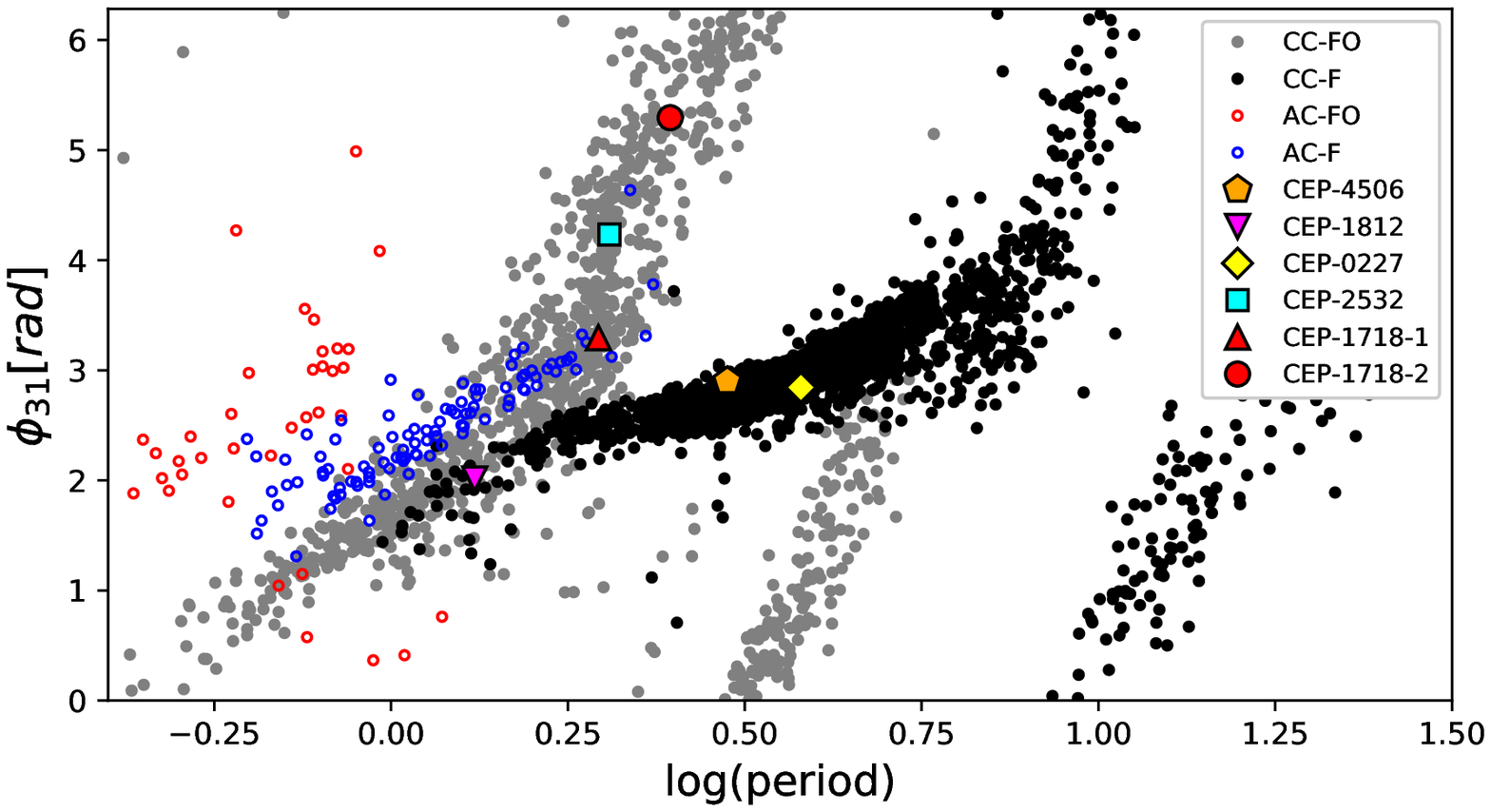}
\caption{Dependence of the light curve shape on pulsation period for fundamental and first-overtone mode classical Cepheids. Selected Fourier decomposition coefficients are shown. The position of the LMC-CEP-1718 components clearly indicates that both Cepheids pulsate in the first-overtone mode. Although not far away from F-mode Anomalous Cepheids, the position of LMC-CEP-1812 is more consistent with classical Cepheids.
\label{fig:all_fourier}}
\end{figure}

In the light curve modeling we have used both the calculated surface brightness value and the radii ratio.  This step introduces quite a significant uncertainty, as we are not relying on the measurements for the given object, but on the statistics of similar objects, so the assumption that they are of the same type is necessary. But as the pulsational light curves of the components are typical for first-overtone classical Cepheids (see Fig.~\ref{fig:all_fourier}) and the sum of the light of the both Cepheids matches the observed light of the system very well, we conclude that this is a safe assumption. Moreover, an alternative solution without these constraints is very unrealistic, resulting in the position of one Cepheid significantly below and the other significantly above the P-L relation for FO classical Cepheids.

\begin{figure}
\centering
\includegraphics[width=0.62\textwidth]{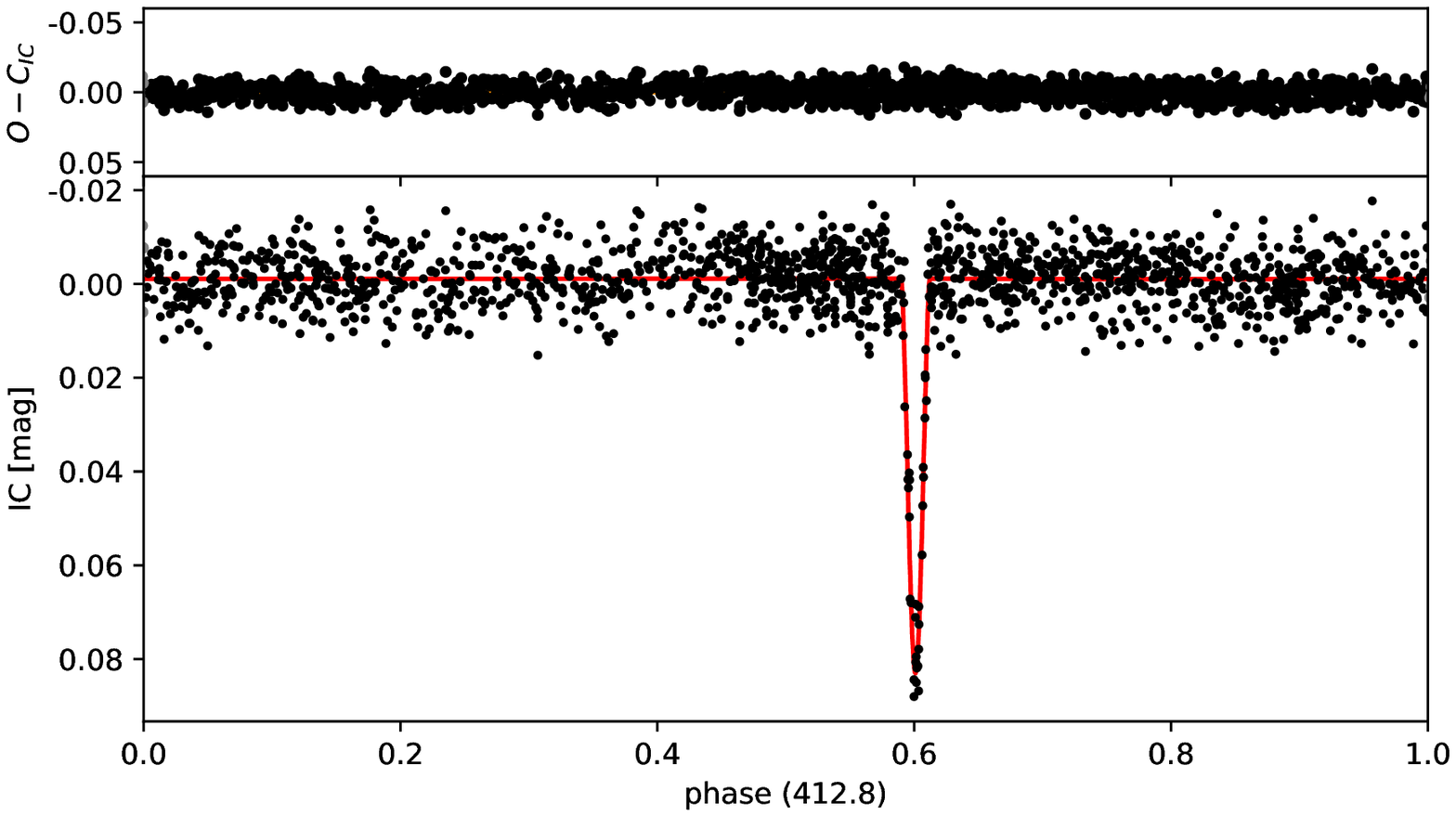}
\includegraphics[width=0.355\textwidth]{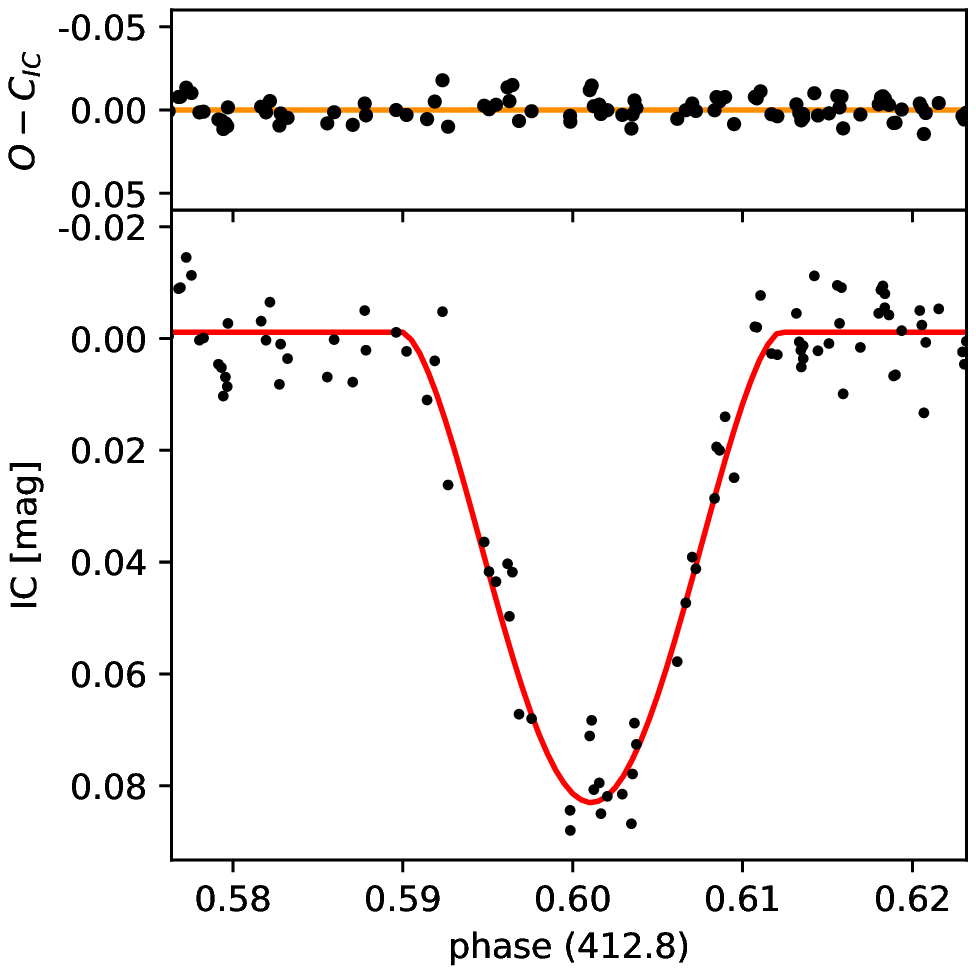}
\caption{Light curve model for LMC-CEP-1718. Contrary to the radial velocity plot, here phase 0.0 corresponds to the superior conjunction, i.e. the phase at which the primary eclipse should occur if the system were more inclined. The right panel shows an expanded view of the secondary eclipse.
\label{fig:1718_lcmodel}}
\end{figure}

The parameters obtained from this new light curve model (see Fig.~\ref{fig:1718_lcmodel}) differ slightly from the ones obtained in \citet{cep1718apj2014}. The inclination is significantly lower and the sum of the relative radii ($0.128 \pm 0.003$) significantly higher than in the previous solution. The time of the secondary eclipse was also calculated from the model and then used to obtain the orbital solution in an iterative way.

\begin{figure}
\begin{center}
  \resizebox{0.6\linewidth}{!}{\includegraphics{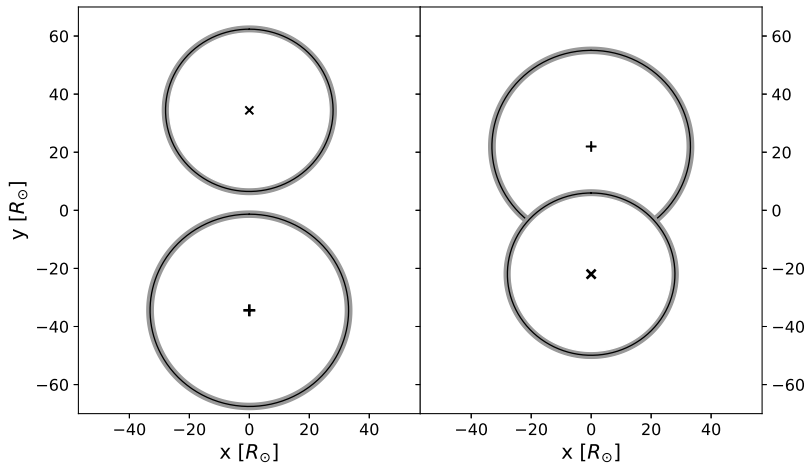}}
\caption{Orbital configuration at the superior and inferior conjunctions for LMC-CEP-1718. The secondary is eclipsed at the inferior conjunction. Gray area marks 1-$\sigma$ errors of the radii.
\label{fig:1718_stars2}}
\end{center}
\end{figure}

Eventually, we calculated the physical parameters using both the orbital solution and the light curve model. They are presented in Table~\ref{tab:1718abs}. The derived masses are similar to but slightly higher than the masses of the other Cepheids in binary systems we have analyzed so far. The radii of the stars are also close to what we expect for FO classical Cepheids of these periods (or rather their sums).
From the theoretical radii for FO Cepheids provided by \citet{Bono2001ApJ...552L.141B} we obtain $64.9 \pm 0.9 R_\odot$ and $59.5 \pm 0.8 R_\odot$ for their canonical and non-canonical models, respectively. Our sum $60.9 \pm 1.5 R_\odot$ falls between both, but is in better agreement with the non-canonical ones. In these models convective core-overshooting during hydrogen-burning phases was taken into account.
The sum of the radii obtained from \citet{2002AstL...28..589S}, i.e.  $57 \pm 5 R_\odot$, is lower but consistent with our value, within the errors.
The configuration of the system during the superior and inferior conjunctions is presented in Fig.~\ref{fig:1718_stars2}. 

Using these results we can try to draw a possible scenario of the  evolutionary state of the components of LMC-CEP-1718. The evolution of the higher mass primary should be more advanced than that of the secondary. However, despite the  very similar temperatures, the secondary is more luminous than the primary, meaning that the evolution of the secondary also has to be more advanced. The mass difference is small enough that this inconsistency might be explained by additional processes, such as mass loss, convective overshooting, or rotation.
In principal, the period change measurements may be used as an indicator of the evolutionary state. In this case the measured values could be interpreted as indicating that the primary is located close to the turning point of the blue loop, where the radius is the smallest, while the secondary is moving blueward on the blue loop.

This explanation is consistent with the characteristics of the primary: with its small radius (as compared to other Cepheids)  it may indeed be located just before or at the blue extreme of the loop. In the case of the secondary the situation is much more complicated. The period decrease and the lower mass are inconsistent with its current position on the HR diagram.
However, random period fluctuations on the time scale  of the OGLE-III and OGLE-IV observations are common among the Cepheids in the LMC \citep{2008AcA....58..313P}. It is then reasonable to conclude that the evolution of the secondary  is  more advanced and that even though it is slightly less massive,  it has already passed through the turning point of the blue loop.

\begin{deluxetable}{lccc}
\tablecaption{Properties of OGLE LMC-CEP-1718 \label{tab:1718abs}}
\tablewidth{0pt}
\tablehead{
\colhead{Parameter} & \colhead{Primary} & \colhead{Secondary} & \colhead{Unit}
}
\startdata
spectral type        & F5 II             & F6 II/Ib          &       \\
pulsation period     & 1.9636625         & 2.480917          & days \\
mass                 & 4.27 $\pm$ 0.04   & 4.22 $\pm$ 0.04   & $M_\odot$ \\
radius\tablenotemark{a}  & 27.8 $\pm$ 1.2 & 33.1 $\pm$ 1.3    & $R_\odot$ \\
$\log g$             & 2.18 $\pm$ 0.04   & 2.02 $\pm$ 0.03   & cgs \\ 
temperature          & 6310 $\pm$ 150    & 6270 $\pm$ 160    & K\\  
$\log L/L_\odot$     & 3.04 $\pm$ 0.06   & 3.18 $\pm$ 0.06   & \\
$V$                  & 15.72 $\pm$ 0.03  & 15.74 $\pm$ 0.03 & mag \\
($V-I$)              &  0.51 $\pm$ 0.02  &  0.52 $\pm$ 0.02 & mag \\
orbital period       & \multicolumn{2}{c}{412.813 $\pm$ 0.008 }   & days \\
$T_{sec}$            & \multicolumn{2}{c}{2456701.77 $\pm$ 0.05 }   & days \\
semimajor axis       & \multicolumn{2}{c}{476.1 $\pm$ 1.2}         & $R_\odot$ \\ 
inclination          & \multicolumn{2}{c}{83.0 $\pm$ 0.5}           & degrees \\ 
$R_1+R_2$            & \multicolumn{2}{c}{60.9 $\pm$ 1.5}           & $R_\odot$ \\
$R_2/R_1$\tablenotemark{b} & \multicolumn{2}{c}{1.19 $\pm$ 0.08}          &  \\
E(B-V)               & \multicolumn{2}{c}{$0.125$ $\pm$ 0.015}       & mag \\
\enddata
\tablenotetext{a}{Calculated from the sum and ratio of the radii. }
\tablenotetext{b}{Radii ratio calculated from the period-radius relations for first-overtone Cepheids. }
\end{deluxetable}

\subsection{OGLE LMC-CEP-1812} \label{subsec:cep1812}

LMC-CEP-1812 is the second classical Cepheid that was confirmed to be a member of an eclipsing binary system, and for which the dynamical masses and other physical parameters were determined \citep{cep1812apj2011}. The state of the evolution of the system is unclear, in part because there is  a large mass difference between the stars, yet they  are in a relatively short stage of  common giant phase evolution  which they should have normally entered after 190 and 369 Myr, respectively. \citet{Neilson2015_cep1812_merger} suggested that the system was originally a triple and the current Cepheid is a merger of two lower mass components of an inner binary. They have also proposed that the Cepheid is crossing the instability strip for the first time, which is quite a rare phenomenon as typically only a few percent of the Cepheids should be found at this stage \citep{2012ApJ...760L..18N}. Moreover they suggested that being a merger, the pulsator may be an Anomalous Cepheid (AC), although it would be a very rare (short-lived) example of a massive AC, which could be identified only because of its binarity.
In this case it is not clear if the star should have properties very different from those of classical Cepheids. For example its light curve shape is typical for a classical Cepheid of this period -- see Fig.~\ref{fig:all_fourier}.

The initial analysis of this object  \citep{cep1812apj2011} used an approximate eclipse light curve model which did not include pulsations. In that model, the pulsations were iteratively subtracted. We have reanalyzed this system using the approach of \citet{cep227mnras2013} for the Cepheid LMC-CEP-0227. Because of its complexity the full analysis of this system will be presented in a separate paper. Here we will only outline the modeling and present the main results.

We have started with the analysis of the periodicity. From the most numerous I-band data (OGLE+EROS) we have measured a  period $P_{IC} = 1.31290385(22)$, while from the longest timespan V-band data (MACHO+OGLE), $P_{VJ} = 1.31290372(15)$, assuming  $\dot{P}=0$.
For the same data sets we have also determined  period changes $\dot{P}_{IC} = (1.1 \pm 1.8) \times 10^{-10}$ and $\dot{P}_{VJ} = (3.2 \pm 1.4) \times 10^{-10}$.
Note that the average observation date for the I-band data set falls 2000 days after the V-band data, and the difference between $P_{IC}$ and $P_{VJ}$ (although within uncertainty limits) can be as well an effect of a positive period change of the order of $10^{-10}$.
Although not statistically significant individually, these three observables together indicate that the pulsation period of LMC-CEP-1812 may be indeed increasing, which would be also consistent with the current evolutionary picture in which the star is crossing the Hertzsprung gap, moving redward on the HR diagram, with its size increasing correspondingly (see also Section~\ref{subsec:evol}).
However, for the purposes of this analysis we can safely assume that the period of the Cepheid is not changing.

To improve the orbital solution we have reanalyzed all the collected spectra, including 9 new observations. In total, we have used 73 spectra (32 MIKE, 28 UVES, 13 HARPS) and have  measured 73 radial velocities for the Cepheid and 68 for the companion (not all of the spectra were of good enough quality and/or taken at good phases to measure a secondary velocity). Instrumental shifts were also fitted using HARPS spectra as a zero-point. We obtained a -0.5 km/s shift for MIKE and +0.55 km/s for the UVES.

\begin{figure}
\centering
\includegraphics[width=0.49\textwidth]{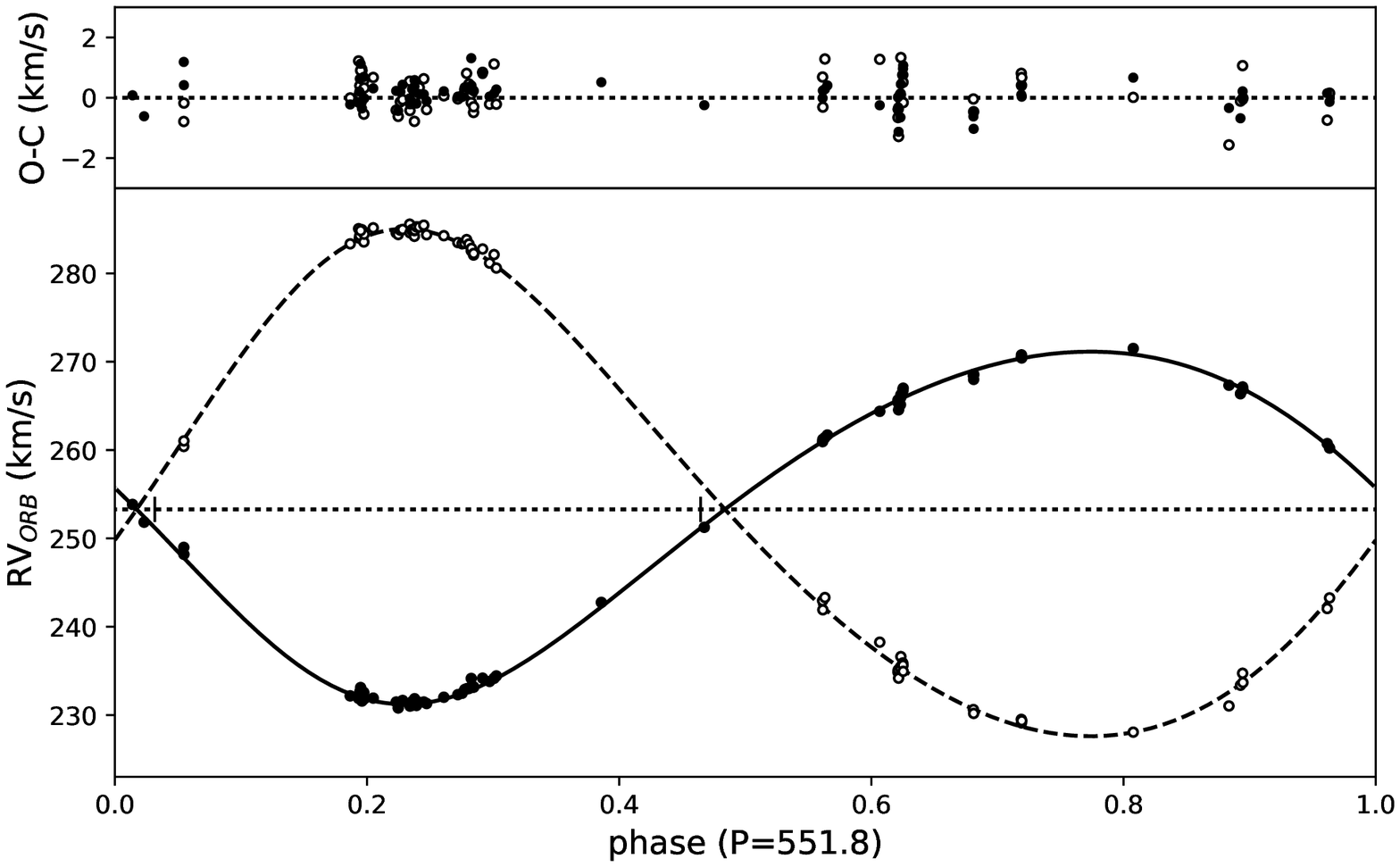}
\includegraphics[width=0.48\textwidth]{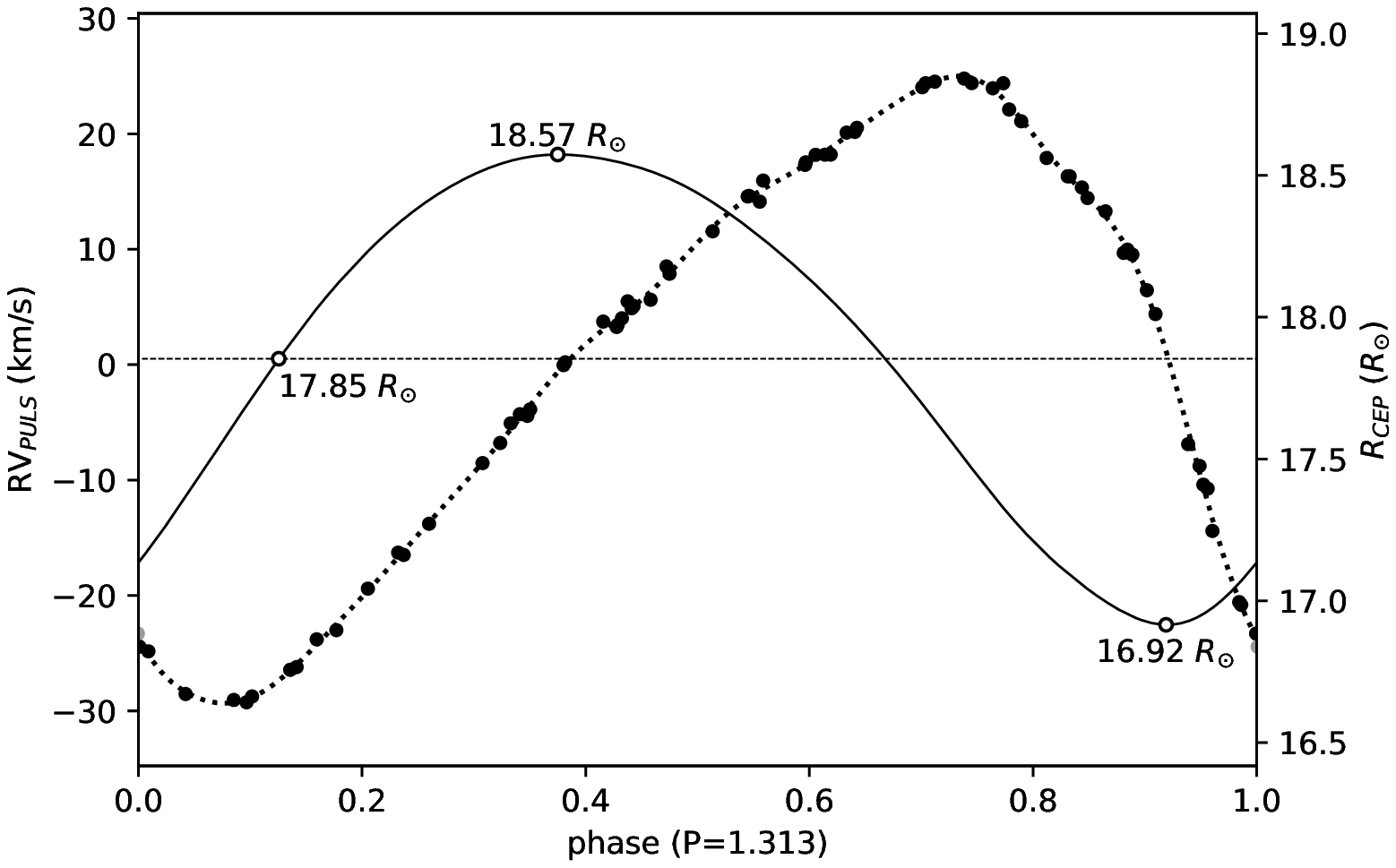}
\caption{{\em left:} Orbital solution for LMC-CEP-1812 with the measured radial velocities of the Cepheid with the pulsations removed (filled circles) and of its companion (open circles). Short vertical lines mark the eclipse phases. {\em right:} Pulsational radial velocity curve (points) and the radius variation of the Cepheid over one pulsation cycle (solid line).
\label{fig:1812_orbpuls}}
\end{figure}

The reference time of phase zero is calculated from the time of the primary minimum, while the eccentricity ($e$) and the argument of periastron ($\omega$) were taken from the photometric solution. For comparison we have also derived $e=0.125$ and $\omega=147.6$ directly from the orbital solution fixing only the times of the eclipses. Although these values are different by about 1.5 $\sigma$ from the photometric solution, this has marginal influence (less than 0.5 $\sigma$) on the derived physical parameters.  We have adopted the values of $e$ and $\omega$ from the photometric solution. The orbital solution and the pulsational radial velocity curve are presented in Fig.~\ref{fig:1812_orbpuls} and the orbital parameters in Table~\ref{tab:1812spec}.

The difference between the $\gamma$ velocities of the components of this system is ($-0.16 \pm 0.12$ km/s) and is much smaller than for the other Cepheids in eclipsing binaries presented in this study.  The small value here may be related to the lowest radius (or highest $\log g$) among the sample.

\begin{deluxetable}{lr@{ $\pm$ }lc}
\tablecaption{Orbital solution for OGLE LMC-CEP-1812  \label{tab:1812spec}}
\tablewidth{0pt}
\tablehead{
\colhead{Parameter} & \multicolumn{2}{c}{Value} & \colhead{Unit}
}
\startdata
$\gamma_1$      & 253.24     &  0.12   &  km/s   \\ 
$\gamma_2$      & 253.39     &  0.08   &  km/s   \\ 
$T_0$ (HJD)     & \multicolumn2c{2455427.49565$^a$} &  d      \\
$a \sin i$      &  525.21    &  1.1    &  $R_\odot$ \\ 
$m_1 \sin^3 i$  &    3.76    &  0.03   &  $M_\odot$ \\ 
$m_2 \sin^3 i$  &    2.62    &  0.02   &  $M_\odot$ \\ 
$q=m_2/m_1$     &    0.696   &  0.003  &  $M_\odot$  \\ 
$e$             & \multicolumn2c{0.1214$^a$} &  -      \\
$\omega$        & \multicolumn2c{150.43$^a$} &  deg  \\
$K_1$           &   19.89    &  0.05   &  km/s \\   
$K_2$           &   28.61    &  0.10   &  km/s \\   
rms$_1$         & \multicolumn{2}{c}{0.48} &  km/s \\
rms$_2$         & \multicolumn{2}{c}{0.67} &  km/s \\
\enddata
 \tablecomments{ $T_0$ is calculated from the epoch of the primary minimum: 2455445.2067.}
 \tablenotetext{a}{Values taken from the photometric solution.}
\end{deluxetable}

In the previous study only the OGLE I-band light curve was modeled. This time we have not only used a more advanced method, but also extended the data by the OGLE V-band and MACHO V and R-band photometric time series, and thus we expect the results to be more reliable and accurate.

\begin{figure}
\begin{center}
  \resizebox{0.55\linewidth}{!}{\includegraphics{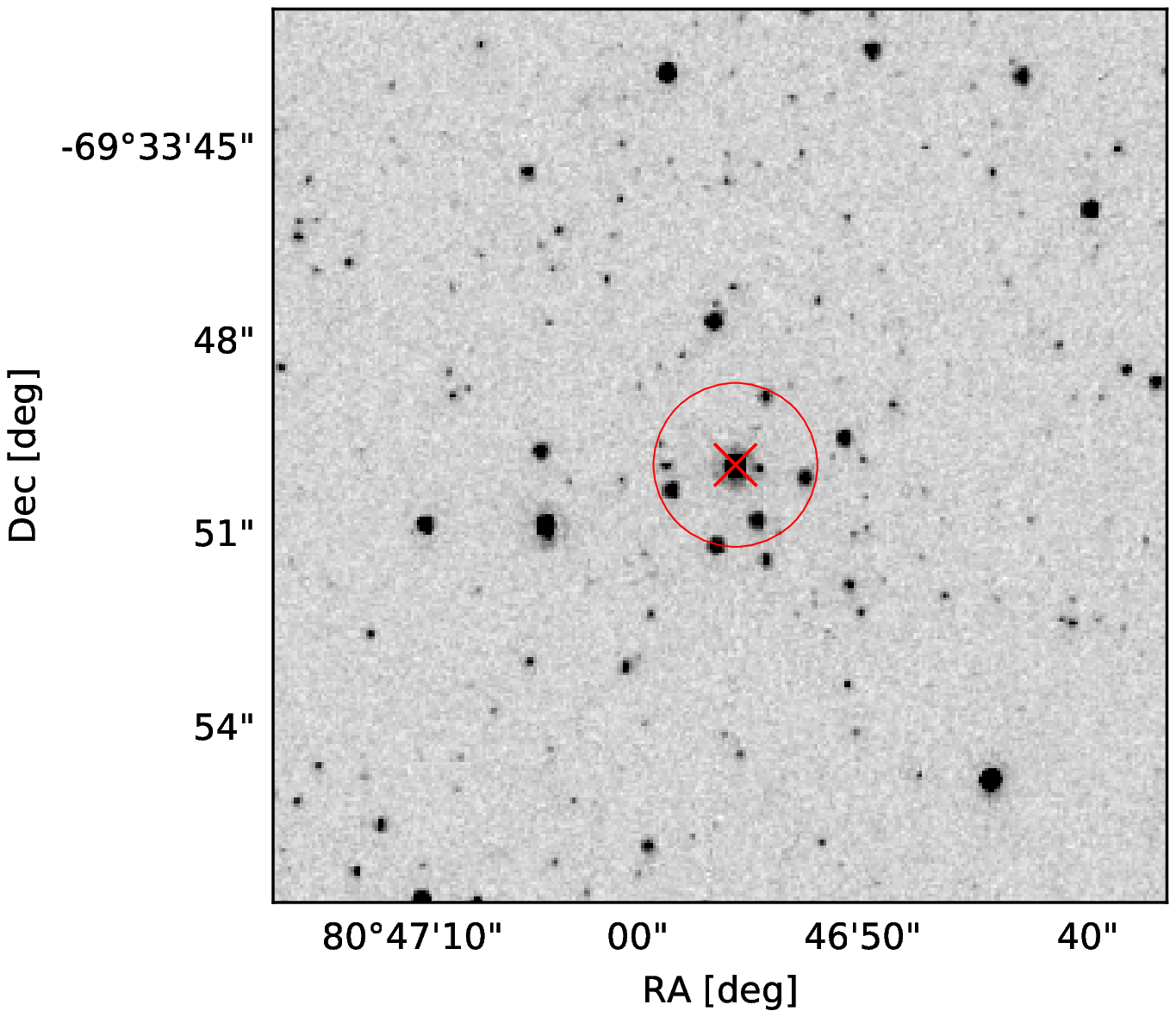}}
  \resizebox{0.398\linewidth}{!}{\includegraphics{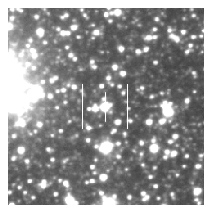}}
\caption{{\em left:} Hubble Space Telescope imaging (program nr. 13010, led by F. Bresolin) of the close neighborhood of the LMC-CEP-1812. The Cepheid is marked with a red cross. There are about seven fainter stars nearby that may influence the ground-based photometry (inside the red circle). {\em right:} A sample OGLE image (60$\times$60 arcsec) for comparison. The white square marks the area shown in the left panel.
\label{fig:1812_hubble}}
\end{center}
\end{figure}

From the analysis of images obtained with the Hubble Space Telescope\footnote{Program nr. 13010, led by F. Bresolin.} and initial modeling trials we found that the OGLE-IV I-band photometry had an incorrect base value of the flux\footnote{The details of the photometric analysis will be presented in a separate paper.}, which was probably related to crowding (see Fig.~\ref{fig:1812_hubble}). To reconstruct the eclipse depths, a very strong negative third light would have to be applied, as the derived inclination is already very close to $90^{\circ}$. We decided to shift the OGLE-IV and OGLE-II data both in flux and magnitude to fit the OGLE-III data. Moreover all the light curves (including the MACHO data) were corrected for long-term brightness variations (including an annual variation). We have also detected cyclical phase shifts related to the orbital motion. This effect (known also as a light-travel time effect or LTTE) was taken into account in the model.

Apart from the new modeling approach, the largest difference between the earlier and present models is  the  determination of the contribution of third light. In the previous study third light was assumed to contribute 10\% of the system light in the I-band as estimated from  the spectra. 
In the current analysis the third light is a free parameter fitted during the modeling.
In spite of the  highly crowded neighborhood, the OGLE-III photometry (to which we transformed the OGLE-II and OGLE-IV data) showed a relatively low additional light of about 2-3 \% depending on the band, i.e. much less than assumed before. On the other hand, the MACHO light curves suffered significantly from the blending with nearby stars -- we detected about 8\% and 15\% of the third light in their R and V bands, respectively. It shows that the previous assumption for the extra light was justified and only the quality of the OGLE-III photometry and flux calibration made this contribution lower than expected.

\begin{figure}
\begin{center}
  \resizebox{\linewidth}{!}{\includegraphics{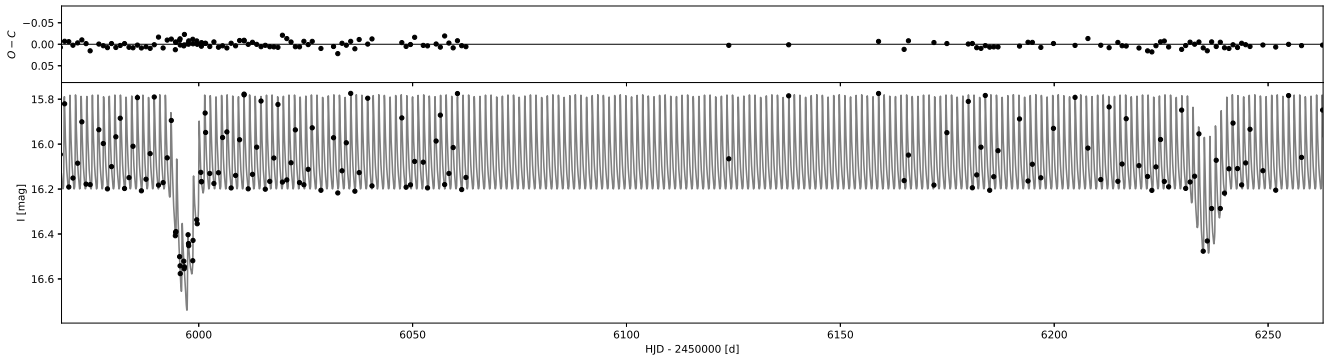}} \\
\caption{I-band model for the LMC-CEP-1812 system. The two eclipses with the highest number of observations are shown.
\label{fig:1812_ic_model}}
\end{center}
\end{figure}

The final light curve model is shown in Fig.~\ref{fig:1812_ic_model}.  In the new solution the precision and accuracy of the measured parameters values was significantly improved and a new geometric p-factor value was determined. We have now three classical Cepheids with the projection factor measured in this way. The value $p = 1.26 \pm 0.08$ is consistent with the values obtained with the interferometry and the SPIPS method \citep{2005A&A...438L...9M,Breitfelder2015_K_Pav,Breitfelder2016_9_CCEP,Kervella2017_RS_Pup,2017A&A...608A..18G}, as well as with theoretical predictions \citep{2009A&A...502..951N}. The value lies between our other measurements of 1.21 for LMC-CEP-0227 and 1.37 for LMC-CEP-4506 (for details see the following sections). It is also in perfect agreement with the value obtained in 2 dimensional simulations of a Cepheid of period 2.8 days \citep{Vasilyev2017_2dcep_model_A&A.606.140}.

In general, in the light curve modeling the third light is correlated with the inclination and the sum of the radii as  these parameters are sensitive to the depth of the eclipses. The lower adopted value of third light (deeper eclipses) is compensated by a lower inclination (shallower eclipses), while  the final sum of the radii has not changed significantly.
However, the change of the ratio of the radii is significant, increasing the derived radius of  the Cepheid by 2.6\% and decreasing the radius of the companion by 2.3\%. The masses follow the same trend, although the effect is much smaller. The properties of the system and the physical parameters of the components are presented in Table~\ref{tab:1812abs}.

\begin{deluxetable}{lccc}
\tablecaption{Properties of OGLE LMC-CEP-1812 \label{tab:1812abs}}
\tablewidth{0pt}
\tablehead{
\colhead{Parameter} & \colhead{Primary} & \colhead{Secondary} & \colhead{Unit}
}
\startdata
spectral type        & F7 III                  & G4 III             &       \\
pulsation period     & 1.31290380(22)          &                    & days \\
projection factor    & 1.26 $\pm$ 0.08         &                    &      \\
mass                 & 3.76 $\pm$ 0.03         & 2.62  $\pm$ 0.02    & $M_\odot$ \\ 
radius               & 17.85 $\pm$ 0.13        & 11.83 $\pm$ 0.08   & $R_\odot$ \\ 
$\log g$             & 2.509 $\pm$ 0.007       & 2.709 $\pm$ 0.007  & cgs \\ 
temperature          & 6120 $\pm$ 150          & 5170  $\pm$ 120     & K \\ 
$\log (L/L_\odot)$   & 2.61 $\pm$ 0.04         & 1.95  $\pm$ 0.04   & \\
$V$                  & 16.87 $\pm$ 0.07      & 18.61 $\pm$ 0.05   & mag \\
($V-I$)              &  0.57 $\pm$ 0.03      &  0.88 $\pm$ 0.03   & mag \\
E($B-V$)             & \multicolumn{2}{c}{$0.04 \pm 0.02$}        & mag \\
eccentricity         & \multicolumn{2}{c}{0.1214 $\pm$ 0.0023 }   &  \\
$\omega$             & \multicolumn{2}{c}{150.4  $\pm$ 1.9 }      & degrees \\
orbital period       & \multicolumn{2}{c}{551.776 $\pm$ 0.003}    & days \\
$T_{I}$              & \multicolumn{2}{c}{5445.207 $\pm$ 0.007 }  & days \\
semimajor axis       & \multicolumn{2}{c}{525.2 $\pm$ 1.1}        & $R_\odot$ \\ 
inclination          & \multicolumn{2}{c}{89.58 $\pm$ 0.04}       & degrees \\
\enddata
\tablecomments{ For the Cepheid the radius, gravity ($\log g$) and the temperature are mean values over the pulsation cycle.}
\end{deluxetable}

From the empirical period-radius relation of \citet{1999ApJ...512..553G} we estimate the radius of the Cepheid to be $16.8 \pm 1.0 R_\odot$. This value is consistent with the measured radius within 1 $\sigma$, although this relation was derived for Cepheids with periods longer than 4 days, and assuming the value of the p-factor from the relation $p = 1.39 - 0.03 \log P$. There seems now to be a general consensus between the observational and theoretical studies that  p-factor values are typically lower (close to 1.3 on average) although a significant internal scatter is seen \citep{2017A&A...608A..18G}. The dependence on the method used is also likely to play a role \citep{2013IAUS..289..138G}.
If we  correct the radius to account for the lower p-factors we  obtain an even lower value (15.3 $R_\odot$ for p=1.26 and 15.8 $R_\odot$ for p=1.3) which is inconsistent with the observed geometrical radius.

\subsection{OGLE LMC-CEP-2532} \label{subsec:cep2532}

LMC-CEP-2532 is a first-overtone Cepheid (P=2.035 days) in a long-period eclipsing binary system (P=800.4 days). Unfortunately, as in the case of LMC-CEP-1718, due to the eccentricity and the orientation of the orbit we see only one eclipse per orbital cycle.
Nevertheless, because there is only one pulsating star in the system and the eclipse is deeper (about 0.35 mag) than for LMC-CEP-1718 (0.09 mag), the analysis is much simpler and not as many assumptions are necessary. This makes this system the only one known to host a first-overtone Cepheid for which fully independent physical parameters can be measured.

In the new OGLE photometry there are no new eclipses observed, neither do we have new photometry from other sources for this system. Therefore, we have only analyzed the new spectroscopic data (acquired with MIKE) and used the new orbital solution to calculate the physical properties of the Cepheid. Comparing with the previous analysis we had 8 new radial velocity measurements, while one old measurement was discarded because of bad quality. Although the additional spectroscopic observations are few in number they have an important impact on the solution as they were taken close to the quadratures at the missing pulsation phases. The improved analysis further enhanced the quality of the results.
Instrumental shifts in the radial velocities were estimated by minimizing the rms values in the orbital fits. Shifts of 500 m/s (HARPS), -600 m/s (MIKE), 350 m/s (UVES) were applied to the measurements.

\begin{figure}
\centering
\includegraphics[width=0.49\textwidth]{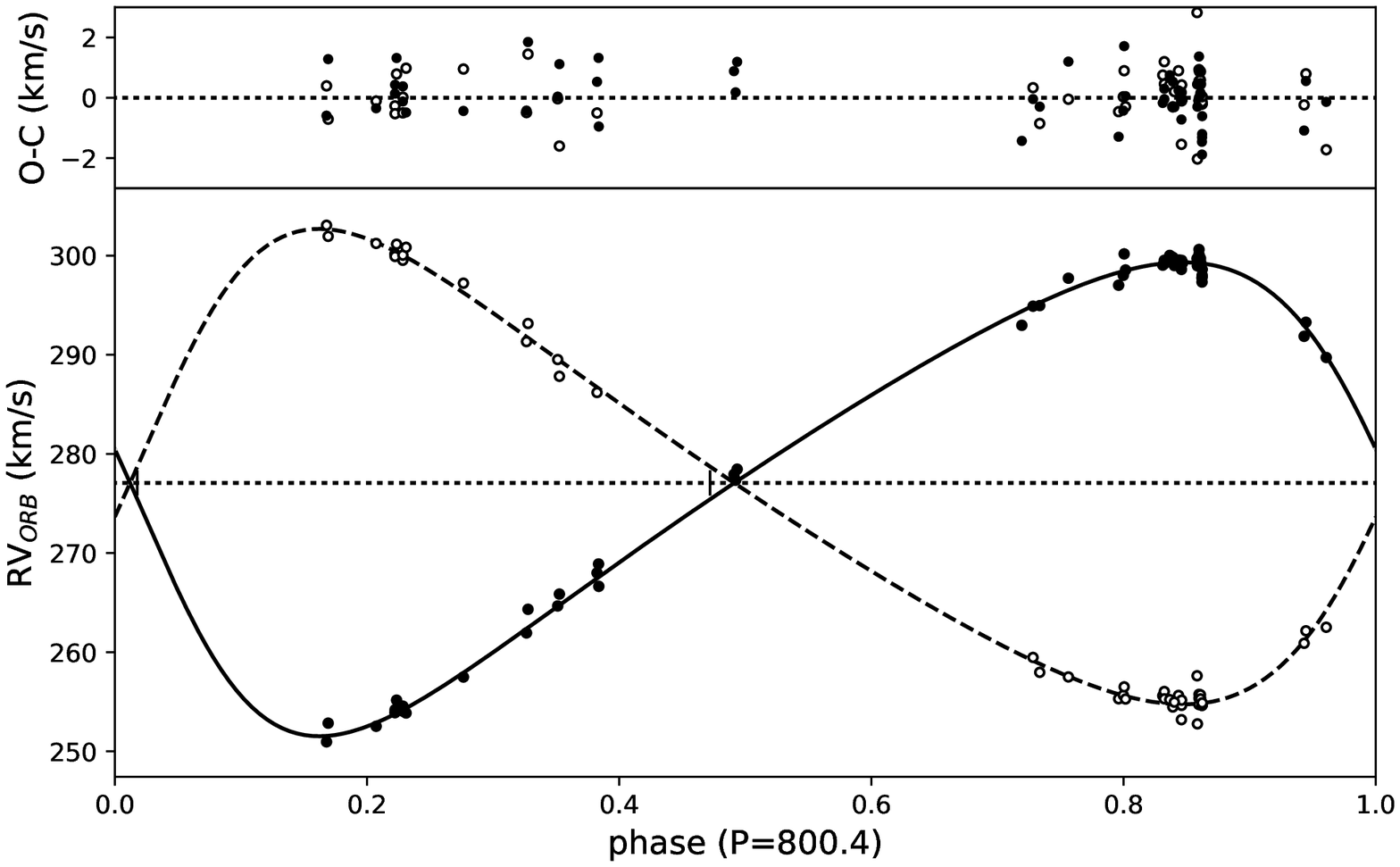}
\includegraphics[width=0.48\textwidth]{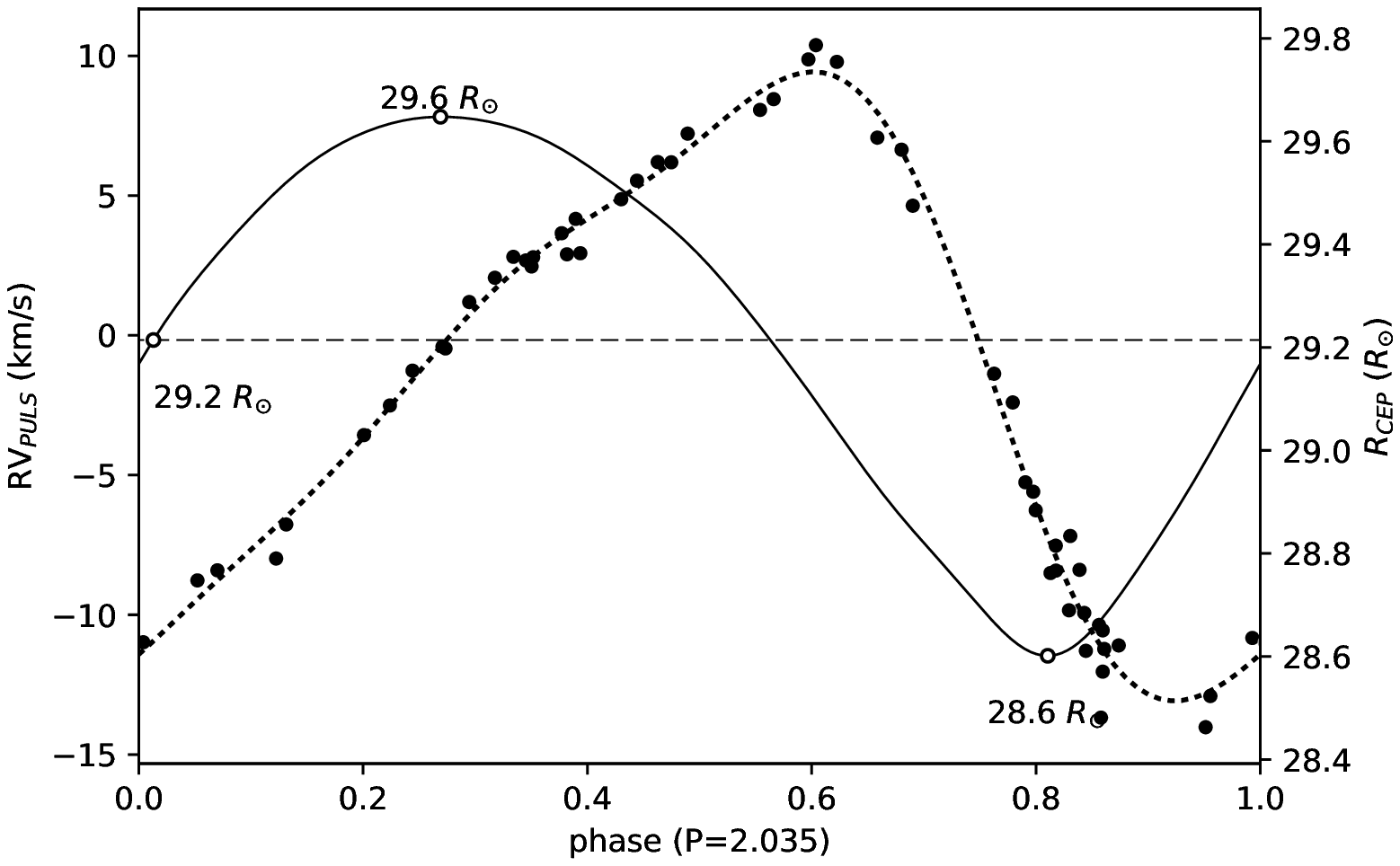}
\caption{Same as Fig.~\ref{fig:1812_orbpuls}, but for LMC-CEP-2532. {\em left:} Short vertical lines mark the conjunction phases. The eclipse (of the Cepheid) is seen in the light curve only at the superior conjunction (thick line). Because of the system configuration there are no eclipses at the inferior conjunction (thin line).
\label{fig:2532_orbpuls}}
\end{figure}

We adopted a 7-order Fourier series to describe the pulsational variation, as higher order fits result in  over-fitting of the pulsation curve. In Fig.\ref{fig:2532_orbpuls} both the orbital and the pulsational radial velocity curves are shown. Note the high eccentricity and the unfortunate orientation of the orbit for which the conjunctions happen close to the lowest and highest separations of the stars (the periastron and apastron phases). 
A large K-term was detected for this Cepheid. The value $-1.3 \pm 0.4$ km/s makes it one of the highest values in our sample.

\begin{deluxetable}{lr@{ $\pm$ }lc}
\tablecaption{Orbital solution for OGLE LMC-CEP-2532  \label{tab:2532spec}}
\tablewidth{0pt}
\tablehead{
\colhead{Parameter} & \multicolumn{2}{c}{Value} & \colhead{Unit}
}
\startdata
$\gamma_1$      &  275.8     &  0.5   &  km/s   \\ 
$\gamma_2$      &  277.05    &  0.23  &  km/s   \\
$T_0$           & \multicolumn{2}{c}{5995.6346 (fixed)} &  d      \\
$a \sin i$      &  721.6     &  5.0   &  $R_\odot$ \\ 
$m_1 \sin^3 i$  &    3.95    &  0.1   &  $M_\odot$ \\ 
$m_2 \sin^3 i$  &    3.92    &  0.09  &  $M_\odot$ \\ 
$q=m_2/m_1$     &    0.992   &  0.012 &  $M_\odot$  \\ 
$e$             &    0.306   &  0.010 &  -      \\
$\omega$        &  104.3     &  2.0   &  deg  \\
$K_1$           &   23.857   &  0.17  &  km/s \\ 
$K_2$           &   24.040   &  0.21  &  km/s \\ 
rms$_1$         & \multicolumn{2}{c}{0.75} &  km/s \\
rms$_2$         & \multicolumn{2}{c}{0.96} &  km/s \\
\enddata
 \tablecomments{ $T_0$ ($HJD - 2450000$~d) is calculated from the epoch of the primary minimum: 5210.8233.}
\end{deluxetable}

The orbital solution is presented in Table~\ref{tab:2532spec}, and the resulting physical properties of the stars in Table~\ref{tab:2532abs}.
The measured radius 29.2 $\pm$ 1.4 $R_\odot$ falls between the theoretical radii for FO Cepheids from \citet{Bono2001ApJ...552L.141B} for canonical ($30.4 \pm 0.4 R_\odot$) and non-canonical ($28.0 \pm 0.4 R_\odot$) models.
The radius is also in good agreement with the radius from the empirical relation of \citet{2002AstL...28..589S}, i.e.  $27 \pm 4 R_\odot$, which is lower but consistent within the errors with the value we have obtained in this study.

\begin{deluxetable}{lccc}
\tablecaption{Physical properties of OGLE LMC-CEP-2532 \label{tab:2532abs}}
\tablewidth{0pt}
\tablehead{
 \colhead{Parameter} & \colhead{Primary (Cepheid)} & \colhead{Secondary} & \colhead{Unit}}
\startdata
spectral type      &  F4 II          & K0 II               & \\ 
pulsational period & 2.03534862      &                     & days \\
mass               & 3.98 $\pm$ 0.10 & 3.94 $\pm$ 0.09     & $M_\odot$\\ 
radius             & 29.2 $\pm$ 1.4  & 38.1 $\pm$ 1.8      & $R_\odot$\\ 
$\log g$           & 2.10 $\pm$ 0.04 & 1.87 $\pm$ 0.04     & {\it cgs}\\ 
temperature        & 6350 $\pm$ 150  & 4800 $\pm$ 220      & K \\ 
$\log L/L_\odot$   & 3.10 $\pm$ 0.06 & 2.84 $\pm$ 0.09     & \\
$V$                & 15.67 $\pm$ 0.05 & 16.44 $\pm$ 0.05   & mag \\
($V-I$)          &  0.50 $\pm$ 0.02 &  1.00 $\pm$ 0.02   & mag \\
E($B-V$)             &\multicolumn{2}{c}{$0.67 \pm 0.08$} & mag \\
orbital period (days) & \multicolumn{2}{c}{800.419 $\pm$ 0.009 } \\
semimajor axis ($R_\odot$) & \multicolumn{2}{c}{723.4 $\pm$ 5.0} \\
\enddata
\tablecomments{ For the Cepheid, the spectral type, radius, gravity ($\log g$), temperature, luminosity ($\log L$) and the dereddened magnitudes and colors are mean values over the pulsation cycle.}
\end{deluxetable}

\subsection{OGLE LMC-CEP-0227} \label{subsec:cep0227}

LMC-CEP-0227 is the first spectroscopically confirmed classical Cepheid in an eclipsing binary system that was analyzed in detail by \citet{cep227mnras2013}. It pulsates in the fundamental mode with a period of 3.797 days and its orbital period is 309.7 days. Because of its  brightness and the almost perfect orientation of the  orbit, this object is perfectly suited for testing pulsation and stellar evolution theories, as the physical parameters of the Cepheid and its companion can be determined with very high precision and accuracy.

Unfortunately, due to a change in observing strategy, the OGLE project photometry after the year 2012 is very scarce. In the I-band there are only 6 points inside the eclipses, while for the V-band there is only 1 measurement. As the inclusion of these data points should not significantly change the solution, we decided to not repeat the light curve modeling and focus only on the radial velocity measurements to improve the orbital solution.

We have obtained a significant number of new spectra with various instruments, 20 with UVES, 3 with MIKE and 13 with HARPS. The total data set consists of  radial velocities for 152 epochs. The new data improve phase coverage around the first quadrature (we had no observations at phases 0.2-0.4 before). We have fitted the instrumental shifts by minimizing the rms values for the orbital fits. We adopted the HARPS velocities   as a reference, and applied shifts of   -650 m/s to the MIKE data and  200 m/s to the UVES data to match the HARPS system.

\begin{figure}
\centering
\includegraphics[width=0.49\textwidth]{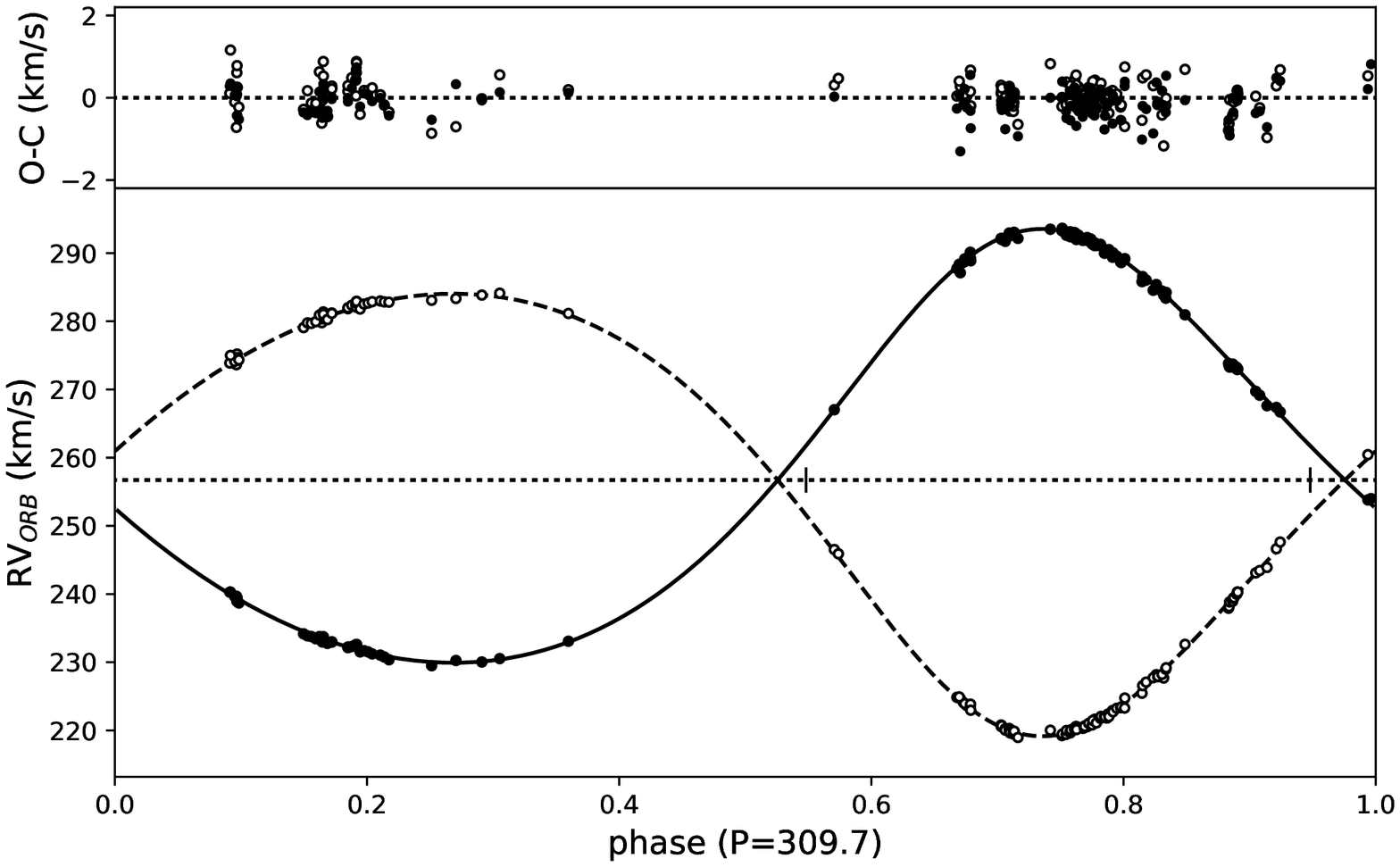}
\includegraphics[width=0.48\textwidth]{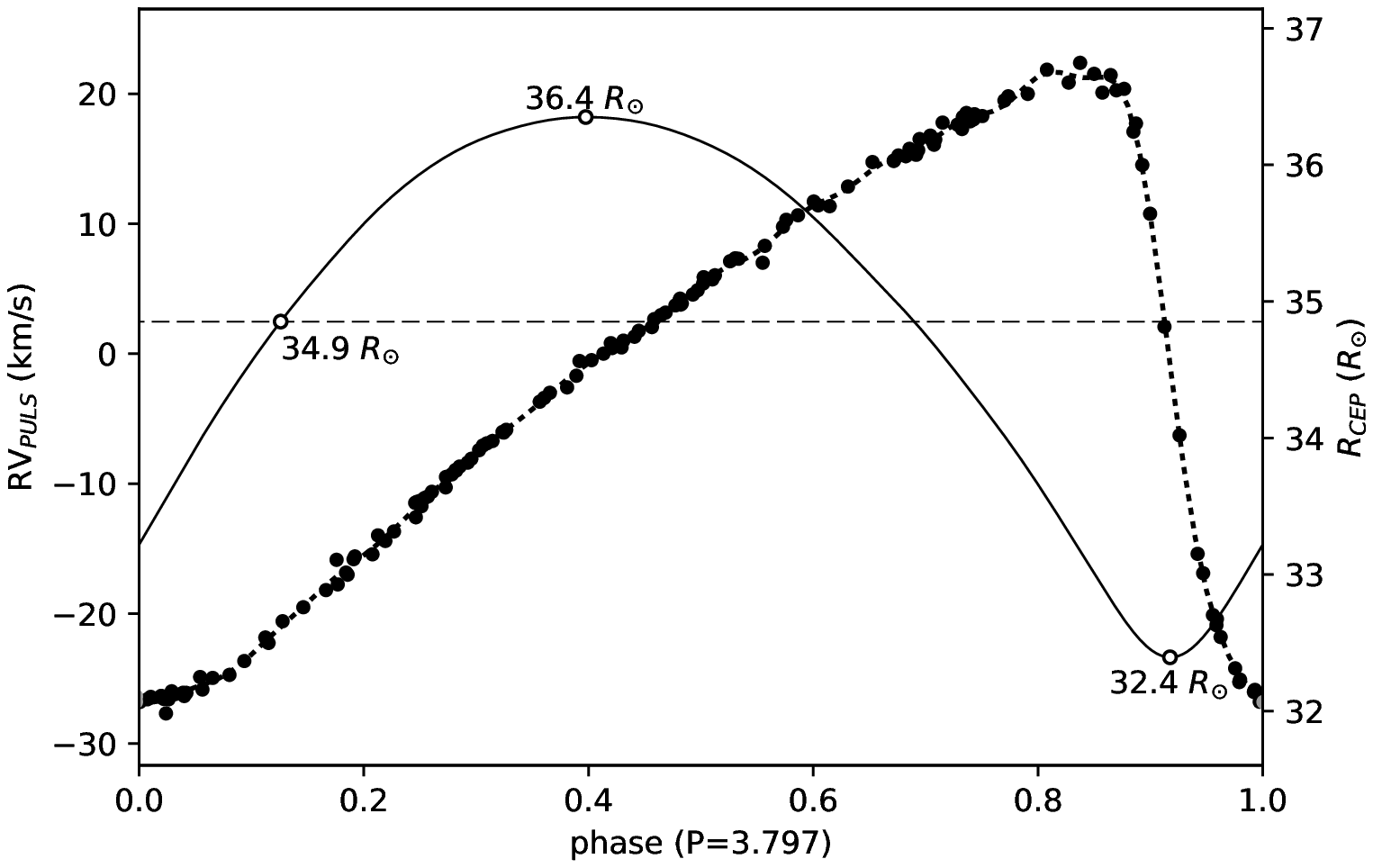}
\caption{Same as Fig.~\ref{fig:1812_orbpuls}, but for LMC-CEP-0227.
\label{fig:0227_orbpuls}}
\end{figure}

The resulting orbital and pulsational radial velocity curves are shown in Fig.\ref{fig:0227_orbpuls}. The new orbital solution is presented in Table~\ref{tab:0227spec}. We have confirmed the significant and strong K-term  \citep{cep227mnras2013}  -- the Cepheid's systemic velocity is blueshifted by $1.12 \pm 0.12$ kms$^{-1}$. This value is not affected by the instrumental shifts, which further enforces the result as a real effect.

\begin{deluxetable}{lr@{ $\pm$ }lc}
\tablecaption{Orbital solution for OGLE LMC-CEP-0227  \label{tab:0227spec}}
\tablewidth{0pt}
\tablehead{
\colhead{Parameter} & \multicolumn{2}{c}{Value} & \colhead{Unit}
}
\startdata
$\gamma_1$      &  255.67     &  0.17  &  km/s   \\
$\gamma_2$      &  256.78     &  0.09  &  km/s   \\
$T_0$           & 5221.68     &  0.16  &  days    \\
$a \sin i$      &  388.2      &  0.8   &  $R_\odot$ \\
$m_1 \sin^3 i$  &    4.13     &  0.02  &  $M_\odot$ \\
$m_2 \sin^3 i$  &    4.04     &  0.03  &  $M_\odot$ \\
$q=m_2/m_1$     &    0.979    &  0.003 & \\
$e$             &   0.1659    &  0.002 &  -         \\  
$\omega$        & 342.0       &  0.7   &  deg  \\       
$K_1$           &   31.799    &  0.06  &  km/s \\ 
$K_2$           &   32.491    &  0.06  &  km/s \\ 
rms$_1$         & \multicolumn{2}{c}{0.40} &  km/s \\
rms$_2$         & \multicolumn{2}{c}{0.39} &  km/s \\
\enddata
\tablecomments{ For the $P_{orb}$ taken from the light curve analysis.}
\end{deluxetable}

The corresponding physical parameters are presented in Table~\ref{tab:0227abs}. As we did not model the system again, the main changes are related to the orbital solution, with the secondary mass being significantly lower than before. All the other parameters are consistent with the previous results. The Cepheid radius is in excellent agreement with the value $34.7 \pm 2.1 R_\odot$ obtained from the empirical period-radius relation of \citet{1999ApJ...512..553G}.

\begin{deluxetable}{lccc}
\tablecaption{Physical properties of OGLE LMC-CEP-0227 \label{tab:0227abs}}
\tablewidth{0pt}
\tablehead{
 \colhead{Parameter} & \colhead{Primary (Cepheid)} & \colhead{Secondary} & \colhead{Unit}}
\startdata
spectral type      &  F7 II/Ib         &  G4 II/Ib         &      \\
pulsational period & 3.797086          &                   & days \\
p-factor           & 1.21  $\pm$ 0.05  &                   &      \\
mass               &  4.15 $\pm$ 0.03  &  4.06 $\pm$ 0.03  & $M_\odot$ \\ 
radius             & 34.87 $\pm$ 0.12  & 44.79 $\pm$ 0.14  & $R_\odot$ \\ 
$\log g$           & 1.970 $\pm$ 0.004 & 1.743 $\pm$ 0.003 & {\it cgs}\\ 
temperature        & 6000  $\pm$ 160   & 5100  $\pm$ 120   & K\\ 
$\log L/L_\odot$    & 3.15  $\pm$ 0.05  & 3.09  $\pm$ 0.04  & \\
$V$                & 15.55  $\pm$ 0.04 & 15.90 $\pm$ 0.05 & mag \\
($V-I$)            &  0.59  $\pm$ 0.05 &  0.88 $\pm$ 0.05 & mag \\
E($B-V$)             &\multicolumn{2}{c}{$0.12 \pm 0.02$}  & mag \\
orbital period (days) & \multicolumn{2}{c}{309.6690 $\pm$ 0.0017 } \\
semimajor axis ($R_\odot$) & \multicolumn{2}{c}{388.8 $\pm$ 0.8} \\  
\enddata
\tablecomments{ The spectral type, radius, gravity ($\log g$), temperature, luminosity ($\log L$) and the dereddened magnitudes and colors are mean values over the pulsation cycle.}
\end{deluxetable}

Evolutionary models have been constructed for LMC-CEP-0227 by various authors. \citet{Cassisi_Salaris_2011ApJ.728L.43} suggested that using extended convective cores (as calibrated independently from the morphology of color-magnitude diagrams of open clusters -- see also \citealt{2004ApJ...612..168P})
and canonical mass loss ($\eta=0.4$) one can obtain excellent consistency with the dynamical mass and explain the evolutionary status of the system.
Similar results were obtained by \citet{Prada2012ApJ...749..108P}. However, in both of these models it was assumed that the companion was more massive and more evolutionary advanced than the Cepheid. According to the new results of this paper, this assumption is no longer justified, and these models should be revisited.

\subsection{OGLE LMC-CEP-4506} \label{subsec:cep9009}

LMC-CEP-4506 is a classical fundamental mode Cepheid in an eclipsing binary system which was known as OGLE LMC562.05.9009, before it was included in the OGLE Collection of Variable stars \citep{2015AcA....65..233S}. It was confirmed to be physically bound in a binary system and analyzed by \citet{cep9009apj2015}.

\begin{figure}
\centering
\includegraphics[width=0.46\textwidth]{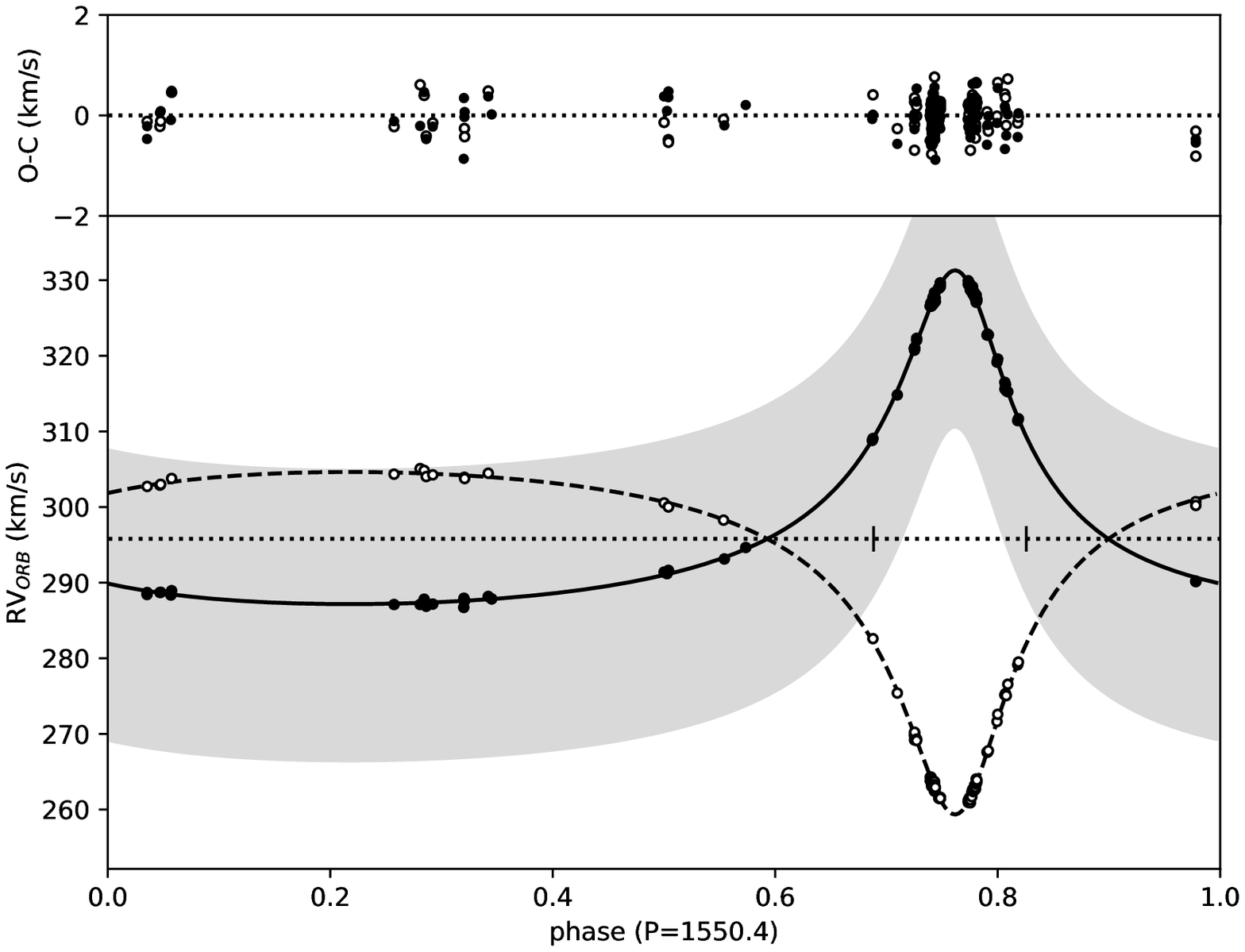}
\includegraphics[width=0.5\textwidth]{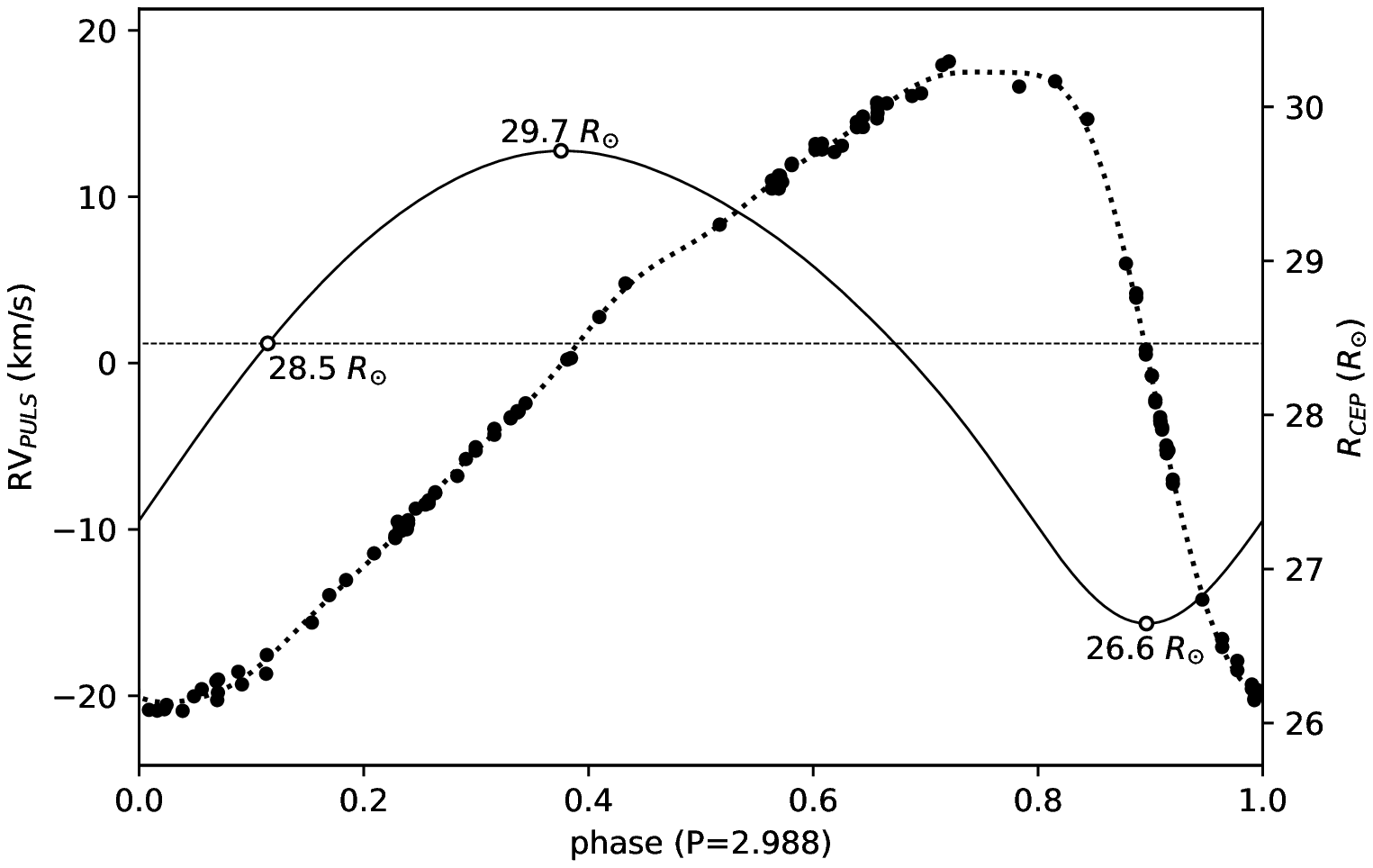}
\caption{Same as Fig.~\ref{fig:1812_orbpuls}, but for LMC-CEP-4506. The gray area marks the velocity range due to the changes in the pulsational radial velocities of the Cepheid. One can see that only for about 10\% of the 4.25-year orbital period (phases 0.7-0.8) the whole pulsation cycle can be observed without problems. 
\label{fig:4506_orbpuls}}
\end{figure}

The binary has a very long orbital period (P=1550d) and very eccentric orbit which makes it difficult to collect well-distributed spectroscopic data. Analysis of the orbit is very challenging, as at  apastron, where the orbital velocities are low, the profiles are blended for a large part of the pulsational cycle -- see Fig.~\ref{fig:4506_orbpuls}. 
Moreover, with a pulsational period of almost exactly 3 days, it is very difficult to achieve uniform coverage of the pulsational RV curve.
The lack of spectroscopic data around the phases corresponding to apastron (0.1 to 0.4) could also in principle affect the mass ratio determination. Fortunately, the orientation of the eccentric orbit breaks the degeneracy between the systemic velocity and the amplitude because of the spiky maximum.

In the current study we have used 38 UVES, 34 MIKE (including 12 new observations) and 10 HARPS spectra. Some of the spectra with blended profiles that were used in \citet{cep9009apj2015} were excluded from the current analysis as they were no longer needed to fill in gaps in the pulsational phases.
In total we have 82 independent data points, compared to 78 used in the previous analysis.

We have used HARPS data as a reference to calculate instrumental shifts in the radial velocities, adopting shifts of -450 m/s (MIKE) and 200 m/s (UVES). The final separated radial velocity curves are presented in Fig.~\ref{fig:4506_orbpuls}. No significant K-term was detected and a single systemic velocity was adopted. We have fitted the orbital parameters fixing the timing of the primary and secondary eclipses taken from the light curve model. The parameters are presented in Table~\ref{tab:4506spec}.

\begin{deluxetable}{lr@{ $\pm$ }lc}
\tablecaption{Orbital solution for OGLE LMC-CEP-4506  \label{tab:4506spec}}
\tablewidth{0pt}
\tablehead{
\colhead{Parameter} & \multicolumn{2}{c}{Value} & \colhead{Unit}
}
\startdata
$\gamma$        &  295.80     &  0.05  &  km/s   \\
$T_0$ (HJD)     & 2457329.8   &  0.9   &  days    \\
$a \sin i$      & 1084.4      &  2.0   &  $R_\odot$ \\ 
$m_1 \sin^3 i$  &    3.60     &  0.02  &  $M_\odot$ \\  
$m_2 \sin^3 i$  &    3.51     &  0.02  &  $M_\odot$ \\  
$q=m_2/m_1$     &    0.975    &  0.003 &  $M_\odot$ \\  
$e$             &    0.6117   & 0.0011 &  -         \\  
$\omega$        &    5.21     &  0.60  &  deg  \\       
$K_1$           &   22.07     &  0.05  &  km/s \\ 
$K_2$           &   22.65     &  0.04  &  km/s \\ 
rms$_1$         & \multicolumn{2}{c}{0.30} &  km/s \\
rms$_2$         & \multicolumn{2}{c}{0.33} &  km/s \\
\enddata
\tablecomments{ For the $P_{orb}$ taken from the light curve analysis.}
\end{deluxetable}

The resulting updated physical parameters are shown in Table~\ref{tab:9009abs}. From the empirical period-radius relation of \citet{1999ApJ...512..553G} we obtain a Cepheid radius of $29.5 \pm 1.8 R_\odot$, which is in very good agreement with our measured value.
As the data for the object come mainly from the OGLE-IV project we have recalculated the reddening using P-L relations based on the OGLE-IV catalog of classical Cepheids. From the recalculated colors of the stars we have obtained a small difference for the temperatures of the components.
The new solution does not solve the problem of the non-pulsating secondary, as discussed in \citet{cep9009apj2015}. Although both stars are located well within the instability strip, only the primary is pulsating. The easiest solution would be that for some reason the temperature is lower than the estimated 6070 K, but a value as low as 5650 K would be necessary, which is quite unlikely, being $2.8-\sigma$ away.

The LMC-CEP-4506 system is a very promising target for future studies. Apart from the observational problems related to the orbital and pulsational period, the brightness of the system  and the deep eclipses make it one of the best examples for studying the structure and evolution of  Cepheids and stars in general.
Observational efforts  should be concentrated during the short duration when the eclipses occur (for photometry) and the orbital velocities are well separated (for spectroscopy).
As we have a good quality light curve for only one orbital cycle at the moment, new photometry will be a crucial ingredient for improving the solution, and especially to reduce the current uncertainty in the determination of the p-factor.

\begin{deluxetable}{lccc}
\tablecaption{Physical properties of OGLE LMC-CEP-4506 \label{tab:9009abs}}
\tablewidth{0pt}
\tablehead{
 \colhead{Parameter} & \colhead{Primary (Cepheid)} & \colhead{Secondary} & \colhead{Unit}}
\startdata
spectral type      & F7 II             &  F7 II            & \\ 
pulsational period & 2.987846(1)       &                   & days \\
p-factor           & 1.35  $\pm$ 0.09  &                   &      \\
mass               &  3.61 $\pm$ 0.03  &  3.52 $\pm$ 0.03  & $M_\odot$ \\ 
radius             & 28.5  $\pm$ 0.2   & 26.4  $\pm$ 0.2   & $R_\odot$ \\
$\log g$           & 2.087 $\pm$ 0.007 & 2.142 $\pm$ 0.006 & {\it cgs} \\
temperature        & 6120  $\pm$ 160   & 6070  $\pm$ 150   & K   \\ 
$\log L/L_\odot$    & 3.01  $\pm$ 0.05  & 2.93  $\pm$ 0.05  &     \\     
$V$                & 15.78  $\pm$ 0.03 & 15.975 $\pm$ 0.04 & mag \\
($V-I$)          &  0.57  $\pm$ 0.04 & 0.58   $\pm$ 0.05 & mag \\
E($B-V$)             &\multicolumn{2}{c}{$0.12 \pm 0.02$}  & mag \\
orbital period     & \multicolumn{2}{c}{1550.354 $\pm$ 0.009 } & days      \\
semimajor axis     & \multicolumn{2}{c}{1085.1   $\pm$ 2.0}      & $R_\odot$ \\ 
inclination        & \multicolumn{2}{c}{  87.98  $\pm$ 0.01}      & degrees   \\
\enddata
\tablecomments{ The spectral type, radius, gravity ($\log g$), temperature, luminosity ($\log L$) and the dereddened magnitudes and colors are mean values over the pulsation cycle.}
\end{deluxetable}

\subsection{OGLE LMC-T2CEP-098} \label{subsec:t2cep098}

This star was the first candidate for a Type II Cepheid in a binary system that was studied by our group, but eventually happened to be an outlier not belonging to any particular Cepheid group, although some similarities with the Anomalous Cepheids were noted.

As a third body influence was detected from the radial velocities of the Cepheid it is important to keep monitoring the system and analyze it again, once the greater part of the outer binary period is covered. It will be also important to have better coverage of the pulsational radial velocity curve to be able to measure the pulsation period more precisely.

Although in principle possible, from the collected data we could not measure the radial velocities of the Cepheid companion. Long-time exposures at carefully selected orbital and pulsational phases should help resolve this problem and let us directly measure the masses, and compare them with the current model.

The physical parameters of our current best model are presented in Table~\ref{tab:t2cep98_abs}. Compared to the work of \citet{t2cep098apj2017} we have used a slightly higher reddening  E(B-V)=0.14, as indicated by a comparison of values obtained from \citet{2011AJ....141..158H} with the values obtained for other Cepheids in the current work.
As LMC-T2CEP-098 does not seem to obey any known period-luminosity relation, we could not use the same method as for other stars directly.

\begin{deluxetable}{lccc}
\tablecaption{Physical properties of OGLE LMC-T2CEP-098 \label{tab:t2cep98_abs}}
\tablewidth{0pt}
\tablehead{
 \colhead{Parameter} & \colhead{Primary (Cepheid)} & \colhead{Secondary} & \colhead{Unit}}
\startdata
spectral type      &  G3 III/II        &     A0 II            &      \\
pulsational period & 4.973726          &                      & days \\
p-factor           & 1.3 $\pm$ 0.05    &                      &      \\ 
mass               &  1.51 $\pm$ 0.09  &   6.8 $\pm$ 0.4      & $M_\odot$\\ 
radius             &  25.3 $\pm$ 0.2   &  26.3 $\pm$ 0.2      & $R_\odot$\\
$\log g$           & 1.81 $\pm$ 0.03   &  2.43 $\pm$ 0.03     & {\it cgs}\\
temperature        & 5300  $\pm$ 120   & 9500  $\pm$ 500      & K        \\ 
$\log L/L_\odot$   & 2.66 $\pm$ 0.04   &  3.71 $\pm$ 0.09     &     \\
$V$                & 16.76 $\pm$ 0.09  & 14.38 $\pm$ 0.09     & mag \\
($V-I$)            & 0.84  $\pm$ 0.04  & -0.02 $\pm$ 0.04     & mag \\
($V-K$)            &  1.8  $\pm$ 0.1   &  0.0  $\pm$ 0.1      & mag \\
E($B-V$)           &\multicolumn{2}{c}{$0.14 \pm 0.03$}       & mag \\
orbital period     & \multicolumn{2}{c}{ 397.178 $\pm$ 0.003} & days \\
semimajor axis     & \multicolumn{2}{c}{460 $\pm$ 6}          & $R_\odot$ \\
\enddata
\tablecomments{ The spectral type, radius, gravity ($\log g$), temperature, luminosity ($\log L$) and the dereddened magnitudes and colors are mean values over the pulsation cycle.}
\end{deluxetable}

\section{Summary and discussion} \label{sec:summary}

We have analyzed five eclipsing binary systems with classical Cepheids and one with a Cepheid of uncertain type 
(LMC-T2CEP-098\footnote{for details of its classification see \citet{t2cep098apj2017}}). Pulsation modes and periods, masses, radii and the p-factor values for these Cepheids are given in Table~\ref{tab:cepdata}. The period range for the fundamental mode Cepheids and the sample size are both too small to obtain well defined relations, but these results allow us to test existing relations. Unfortunately the first-overtone Cepheids are clumped at almost the same period and for all of them only one eclipse per cycle is visible, making the solutions somewhat uncertain. Nevertheless we could obtain some very interesting results for the sample as a whole.

\begin{deluxetable}{lcccccccl}
\tablecaption{Masses, radii and p-factors of Cepheids \label{tab:cepdata}}
\tablewidth{0pt}
\tablehead{
\colhead{ID} & \colhead{Type} & \colhead{Period [d]} & \colhead{P$_F$[d]$^a$} & \colhead{Mass [M$_\odot$]} & \colhead{Radius [R$_\odot$]} & \colhead{A$_{RV}$ [km/s]} & \colhead{p-factor} & \colhead{Comment}
}
\startdata
LMC-CEP-0227   & CC-F  & 3.797086  & -     & 4.15 (3)       & 34.87 (12)    & 48.5 & 1.21 (5) & \\
LMC-CEP-4506   & CC-F  & 2.987846  & -     & 3.61 (3)       & 28.5 (2)      & 38.1 & 1.35 (9) & \\
LMC-CEP-2532   & CC-FO & 2.035349  & 2.833 & 3.98 (10)      & 29.2 (1.4)    & 23.6 &   -      & OEPC \\
LMC-CEP-1718B  & CC-FO & 2.480917  & 3.460 & 4.22 (4)       & 33.1 (1.3)    & 21.5 &   -      & OEPC \\
LMC-CEP-1718A  & CC-FO & 1.963663  & 2.732 & 4.27 (4)       & 27.8 (1.2)    & 17.6 &   -      & OEPC \\
LMC-CEP-1812   & CC-F  & 1.312903  & -     & 3.76 (3)       & 17.85 (13)    & 54.3 & 1.26 (8) & \\
\hline
LMC-T2CEP-098  & other-F$^b$ & 4.97326 & -   & 1.51 (9)     & 25.3 (2)      & 54.9 & 1.30 (5) & SB1 \\
\enddata
\tablecomments{OEPC -- one eclipse per cycle, SB1 -- single-lined spectroscopic binary. Uncertainties in the last one or two digits are given in parentheses.}
\tablenotetext{a}{Fundamental mode period from a pulsation theory model for a given FO Cepheid.}
\tablenotetext{b}{Outlier located between CC and T2C, probably a long-period Anomalous Cepheid.}
\end{deluxetable}

\subsection{Cepheid Period-Luminosity, Period-Radius, and Period-Mass-Radius Relations}

\begin{figure}
\begin{center}
  \resizebox{0.6\linewidth}{!}{\includegraphics{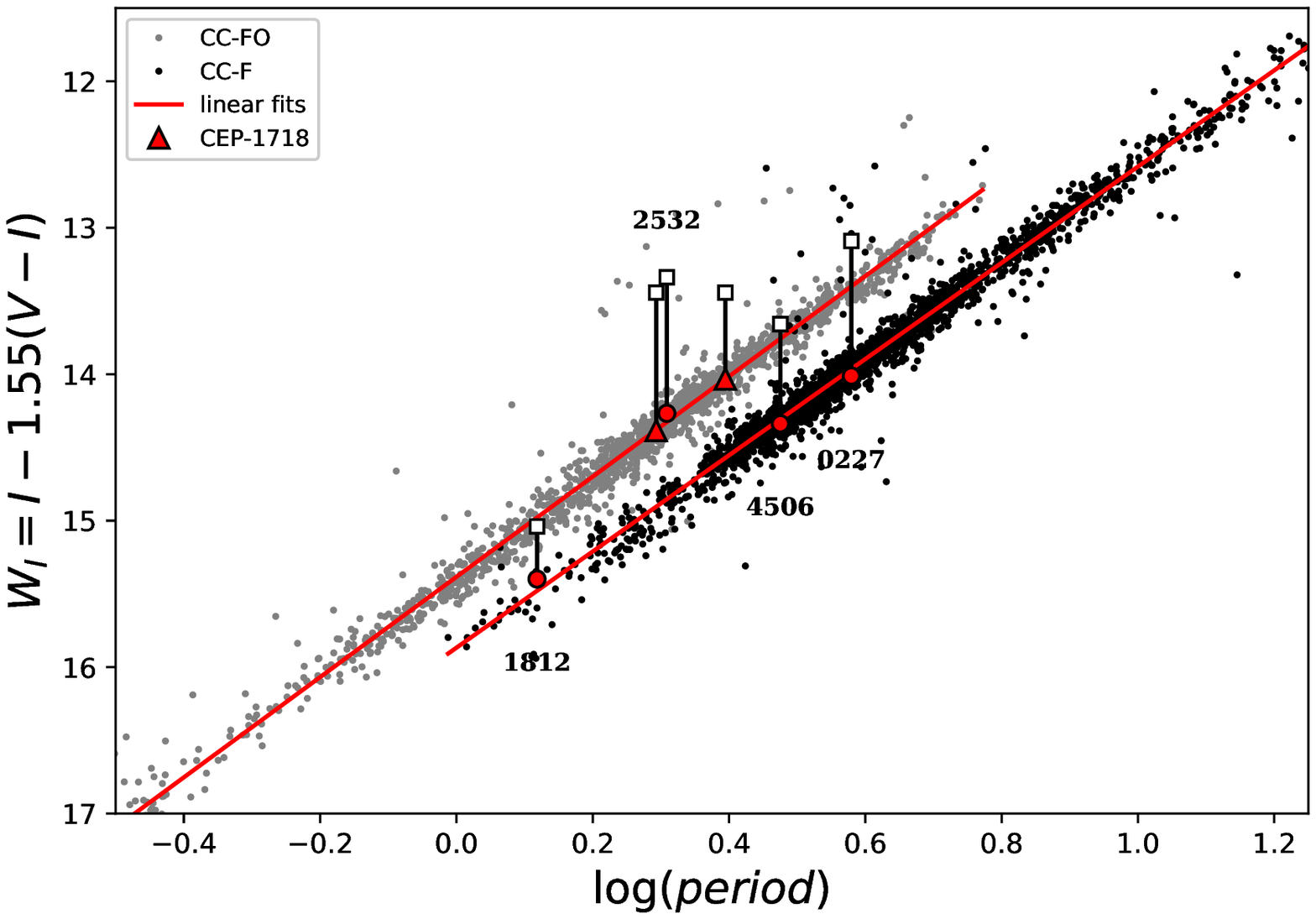}} \\
\caption{Period-luminosity diagram for Classical Cepheids. The Cepheids are marked with red circle or triangle, while the system brightness is marked with a square. Note that for LMC-CEP-1718 the light ratio was taken from the P-L relation itself, but the total light comes from the observations. Linear fits to fundamental and first-overtone sequences are also shown.
\label{fig:all_pl}}
\end{center}
\end{figure}

Our sample of classical Cepheids on the period-luminosity relation for classical Cepheids is shown in Fig.~\ref{fig:all_pl}. This study confirms  previous classification. Although originally lying considerably above the relation (empty squares), after subtraction of the companion light they move onto their respective PL relations (fundamental and first overtone mode). In the case of LMC-CEP-1718 the light ratio was taken from the relation, but the total light is the observed value. As we do not see any vertical shift, our assumption that both components are first-overtone classical Cepheids is confirmed.

\begin{figure}
\begin{center}
  \resizebox{0.7\linewidth}{!}{\includegraphics{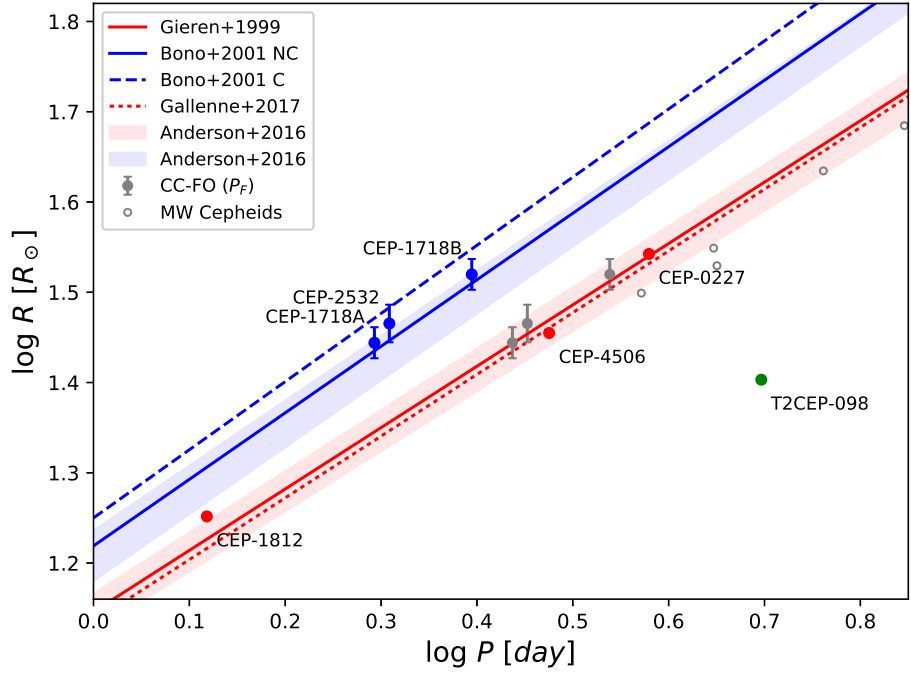}} \\
\caption{Period-radius diagram. The FO Cepheids ({\em blue}) of our study lie between the canonical (C) and non-canonical (NC) theoretical P-R relations, but are more consistent with the latter. The F-mode ({\em red}) Cepheids are consistent with the empirical relation. Color areas mark the theoretical relations with the rotation included for stars crossing the instability strip for the third time. For the FO pulsators their positions for corresponding fundamental mode periods ($P_F$) are shown in gray. Radii of several Galactic Cepheids in this period range are also shown for comparison.
\label{fig:all_pr}}
\end{center}
\end{figure}

In Fig.~\ref{fig:all_pr} we show the position of our sample of Cepheids on the period-radius plane. First-overtone Cepheids lie between the canonical and non-canonical relations of \citet{Bono2001ApJ...552L.141B}. Although the uncertainties are large, our results are more consistent with the latter, indicating  that convective core-overshooting during hydrogen-burning phases should be taken into account.
 We find excellent agreement for the fundamental mode Cepheids, although LMC-CEP-1812 seems to be somewhat over-sized for its period. It is important to note that the relation of \citet{1999ApJ...512..553G} was obtained for a sample of Cepheids with periods longer than 4 days, and we are relying on an extrapolation to this range. We have also a good agreement with a recent relation by \citet{2017A&A...608A..18G}.
The colored areas mark the theoretical relations for stars crossing the instability strip for the third time \citep{Anderson_2016A&A.591.8}. The relations for Z=0.006 and an average initial rotation rate (their Table 5) for Cepheids pulsating in both modes are shown. As we can see, the inclusion of convective core-overshooting is not the only way to improve the models.

We would like to point our here that in general the theoretical studies were focused on higher mass ($>$5 $M_\odot$) and longer period ($>$4 days) Cepheids. Observational data is also biased towards higher radius/period stars. In the case of binary systems, it is more probable that a less massive, smaller star survives the red giant phase, thus shorter period Cepheids are expected to be found preferentially. To take full advantage of the results from the analysis of Cepheids in eclipsing binary systems, theoretical efforts focused on low-mass ($\sim$4 $M_\odot$) classical Cepheids should be undertaken. It would be also valuable to obtain radii of shorter period stars from interferometric observations for comparison.

\begin{figure}
\begin{center}
  \resizebox{0.7\linewidth}{!}{\includegraphics{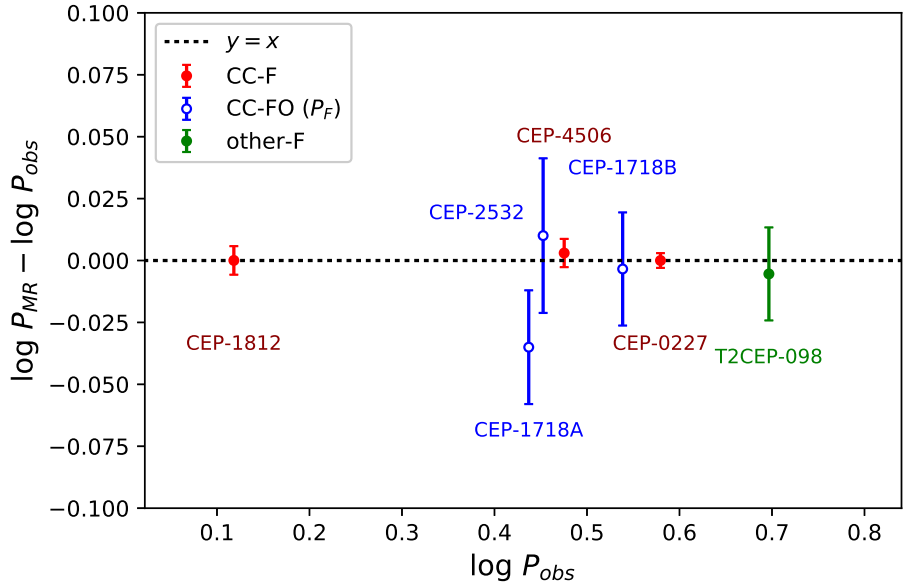}} \\
\caption{ Difference between the period calculated from the PMR relation ($\log P_{MR}$) derived in this paper and the observed period ($\log P_{obs}$). The relation is based mostly on three fundamental mode classical Cepheids, but agrees well with all the pulsating stars in the sample. For the first-overtone pulsators their corresponding fundamental mode periods ($P_F$) were used.
The rms scatter of the relation is 0.037 for the whole sample and 0.003 for fundamental mode Cepheids only.
\label{fig:all_pmr}}
\end{center}
\end{figure}

Using the periods, masses and radii of the Cepheids analyzed in this study, we have fitted a PMR relation in the form $\log P = a + b\log M + c\log R$ and obtained:
$$ \log P_{MR} = -1.555(35) - 0.795(44) \log M + 1.703(23) \log R, $$
where $P_{MR}$, $M$ and $R$ are expressed in days, $M_\odot$ and $R_\odot$, respectively. The rms scatter of the relation is 0.037 for the whole sample and 0.003 when only fundamental mode Cepheids are taken into account. Our fit and the data used for fitting are shown in Fig.~\ref{fig:all_pmr}.  For the first-overtone pulsators, their corresponding fundamental mode periods ($P_F$) were taken as calculated by \citet{2018arXiv180310911T} using the pulsation code of \citet{sm08a} and assuming [Fe/H] = -0.5 dex. They are given in Table~\ref{tab:cepdata}.
The parameters of this relation are mostly defined by the parameters of the three fundamental mode classical Cepheids, which are the best measured and reliable ones, but they do describe well all the other variables in our sample.
Except the intercept coefficient (a), the others (b,c) are consistent with the coefficients of relations given by \citet{Bono2001_MR_ApJ.563.319B} and once again the consistency is moderately better with the non-canonical relation.

The current empirical calibration of the PMR relation will be very important for determining the masses of single Cepheids for which the period is easily measured, and the radius can be determined from interferometric or spectrophotometric observations from a Baade-Wesselink analysis. The tightness of the relation is especially promising, and although more data points for shorter and longer pulsation periods are important to strengthen the relation, it can be used with confidence for a wide range of periods (i.e. about 1.25 to $\sim$10 days).

As an example of the use of the relation in Table~\ref{tab:cepmass} we present masses for Galactic Cepheids for which radii were measured with the SPIPS method by \citet{Breitfelder2015_K_Pav,Breitfelder2016_9_CCEP} and \citet{Kervella2017_RS_Pup}. These radii are systemically lower than those of our LMC Cepheids for their corresponding periods (see Fig.~\ref{fig:all_pr}), which is consistent with the higher metallicity of the Milky Way Cepheids and results in their lower masses.
We note that the calculated mass of the Type II Cepheid $\kappa$ Pav, $0.56 \pm 0.08 M_\odot$, is in very good agreement with the masses expected for this type of variables, i.e. $0.5-0.6 M_\odot$  \citep{2002PASP..114..689W,1997A&A...317..171B,2016CoKon.105..149B}. The values obtained for the two classical Cepheids with the longest periods are strongly extrapolated and should be treated with caution, as both periods and radii are far outside of parameter ranges covered by our sample.

\begin{deluxetable}{lcccccl}
\tablecaption{Masses of Galactic Cepheids \label{tab:cepmass}}
\tablewidth{0pt}
\tablehead{
\colhead{ID} & \colhead{Period [d]} & \colhead{Radius [R$_\odot$]} & \colhead{Mass [M$_\odot$]} & \colhead{p-factor$^a$} & \colhead{Comment}
}
\startdata
$\kappa$ Pav & 9.0827 &  22.8 (1.1) &  0.56 (8) & 1.26 (4,6)  & T2C \\
\hline
RT Aur  & 3.728305  &  31.6 (3.1) &  3.4 (6) & 1.20 (8,9)  & \\
 T Vul  & 4.435424  &  35.4 (5.0)  &  3.5 (6) & 1.48 (4,18) & \\
FF Aql  & 4.470848  &  33.8 (2.7)  &  3.2 (6) & 1.14 (7,7)  & binary\\
 Y Sgr  & 5.773383  &  43.1 (6.7) &  3.9 (7) & 1.31 (6,18) & \\
 X Sgr  & 7.012770  &  48.4 (3.8)  &  3.9 (7) & 1.39 (4,8) & \\
 W Sgr  & 7.594984  &  54.6 (5.6) &  4.5 (8) & 1.35 (6,12) & \\
$\beta$ Dor &  9.842675 & 62.1 (3.6) & 4.3 (8) & 1.36 (4,7) & \\
$\zeta$ Gem & 10.149806 & 64.9 (6.1) & 4.6 (8) & 1.41 (4,9) & \\
\hline
$\mathcal{L}$ Car & 35.55161 & 159 (17) & 6.4$^b$ (1.3) & 1.23 (1,12) & long period \\
           RS Pup & 41.43814 & 191 (22) & 7.9$^b$ (1.7) & 1.25 (3,6) & long period \\ 
\enddata
\tablecomments{Periods, radii and p-factors taken from \citet{Breitfelder2015_K_Pav,Breitfelder2016_9_CCEP} and \citet{Kervella2017_RS_Pup}. Masses calculated using the PMR relation derived in this paper. Uncertainties in the last one or two digits are given in parentheses.}
\tablenotetext{a}{Statistical and systematic errors are given.}
\tablenotetext{b}{Value should be used with caution because of strong extrapolation.}
\end{deluxetable}

\subsection{Evolutionary Status} \label{subsec:evol}

\begin{figure}
\begin{center}
  \resizebox{0.7\linewidth}{!}{\includegraphics{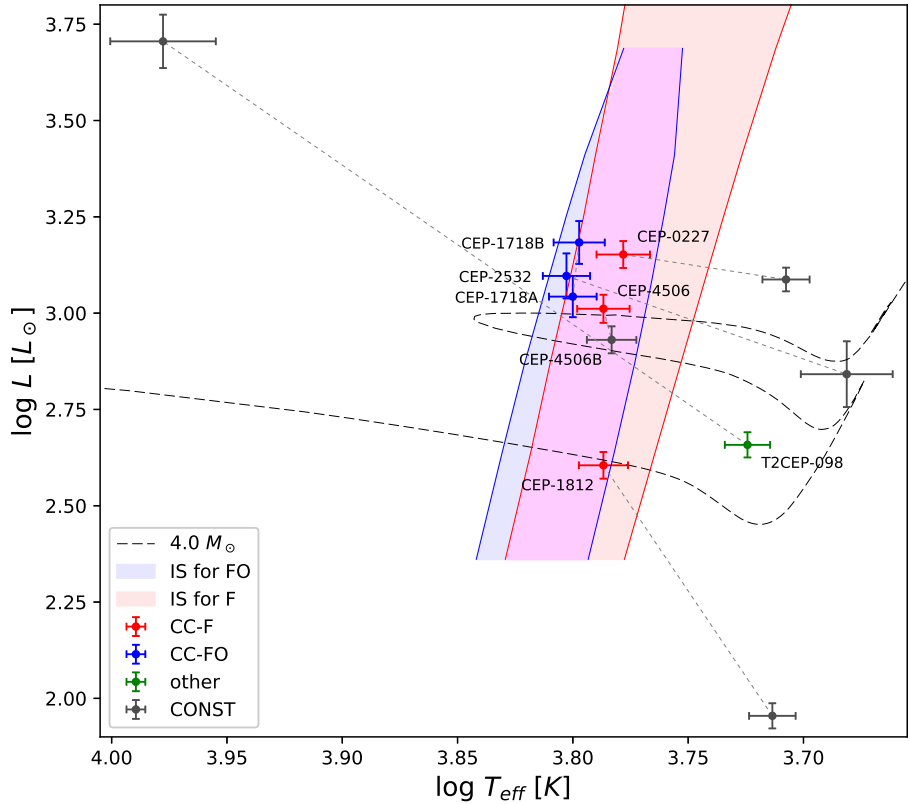}} \\
\caption{Position on the Hertzsprung-Russell diagram for the analyzed Cepheids and their companions. Theoretical boundaries for fundamental (red) and first-overtone (blue) mode classical Cepheids are shown. A sample evolutionary track for a 4 M$_\odot$ star and [Fe/H] $\sim$ -0.5 is also shown as a reference.
\label{fig:all_hr}}
\end{center}
\end{figure}

The position of the Cepheids and their companions on the Hertzsprung-Russell diagram is shown in Fig.~\ref{fig:all_hr}. Edges of the instability strip (IS) were determined from a model grid computed with the linear convective pulsation code of \citet{sm08a}. All models assume [Fe/H]=-0.5 and the mass-luminosity relation derived from the blue-loop evolutionary tracks of \citet{georgy}.

Most of our binary systems have mass ratios very close to unity, which increases significantly the probability of finding their both components inside the IS. In our sample there are two such systems. In the case of the LMC-CEP-1718 system, both components indeed pulsate in the first-overtone mode. However, in the other similar system, LMC-CEP-4506, only the primary pulsates, while its companion (with the suffix B in the plot) is a stable star. This situation can hardly be explained with the observational errors as it resides in the center of the IS, and
 parameters very different from the derived ones would be required to move the star out of the IS and make it stable against pulsations.

In the case of LMC-CEP-0227 and LMC-CEP-2532 the companions are cooler and probably less evolutionary advanced, which is also supported by their masses being slightly lower than those of the primary components. The companion of LMC-T2CEP-098 is much brighter and hotter than the pulsating variable star, while the companion of LMC-CEP-1812 is cooler and much fainter.

An evolutionary track for a star with a mass of 4 M$_\odot$ and metallicity [Fe/H] $\sim$ -0.5 is also shown in the plot as a reference. The position of  LMC-CEP-1812 is clearly distinct from the other Cepheids, confirming the conclusion that it is crossing the instability strip for the first time. The evolutionary state of the LMC-T2CEP-098 system cannot be explained without a binary interaction event in the past, as the apparently more evolutionary advanced pulsating component is 4.5 times less massive than its hot companion.

\subsection{P-Factors for Cepheid Variables}

In our method, the instantaneous radius of the star is a function of the p-factor, and the light curve shape depends on it. Depending on the p-factor, eclipses in the model may start or end earlier or later, and the amplitude of light variation during the eclipses is also affected. It is the only direct and geometric method of determining the projection factor applied so far.

\begin{figure}
\begin{center}
  \resizebox{0.65\linewidth}{!}{\includegraphics{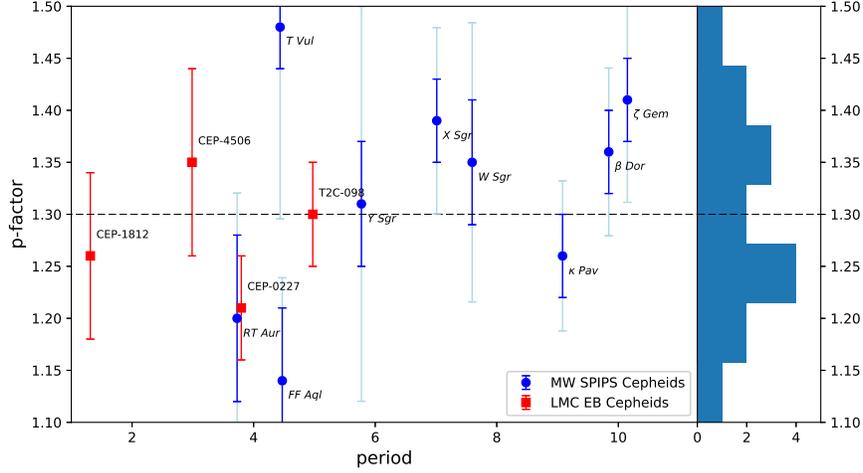}} 
\caption{Period -- p-factor diagram. No correlation is seen neither for our eclipsing binary LMC Cepheids, nor for the Galactic ones. For MW Cepheids, the total uncertainty is shown in a lighter blue color, while the statistical part is marked with a darker one. A histogram for all p-factor values is shown on the right.
\label{fig:pfac-per}}
\end{center}
\end{figure}

In this work we have presented the third measurement of the p-factor value for a classical Cepheid variable in a binary system. All measurements of this factor are presented in Table~\ref{tab:cepdata}. 
Although various period -- p-factor relations have been proposed in the literature \citep{2009A&A...502..951N,Storm_2011_pfac_AA.534.94,2017A&A...608A..18G}, we cannot see any clear corresponding trend in our data. We have to admit however that the period range covered by our sample of Cepheids is relatively small -- see Fig.~\ref{fig:pfac-per}.

Nevertheless, some observations can be made regarding the sample. Comparing the Cepheids with the highest (LMC-CEP-4506) and lowest (LMC-CEP-0227) p-factor we find that although the periods are quite similar, the former has a significantly lower mass and a lower pulsational RV curve amplitude ($A_{RV}$) -- see Table~\ref{tab:cepdata}. The p-factor of LMC-T2CEP-098 is moderately high, while the star has the lowest mass and exhibits the highest RV amplitude. Two stars with the same RV amplitudes (T2CEP-098 and CEP-1812) have quite similar p-factors and anti-correlated masses and periods.
This suggests that the p-factor may depend not only on the period, but perhaps also on RV amplitude and the mass of the pulsating star (the dependence on the latter was already suggested by \citealt{2017EPJWC.15207007P}). Individual fits to only the mass or only the RV amplitude show a clear trend with only one, different outlying star in each case. However, we do not have yet enough data points to fit a two-parameter relation.

We have also checked if any similar relation is seen when the Galactic Cepheids (except the two longest-period ones) from Table~\ref{tab:cepmass} are added. Although the masses are not as precisely determined, the errors are highly correlated (meaning no significant relative shifts between the stars) and as the relation used to obtain them was based on our mass determination, they should be also consistent with our measurements.

\begin{figure}
\begin{center}
  \resizebox{0.49\linewidth}{!}{\includegraphics{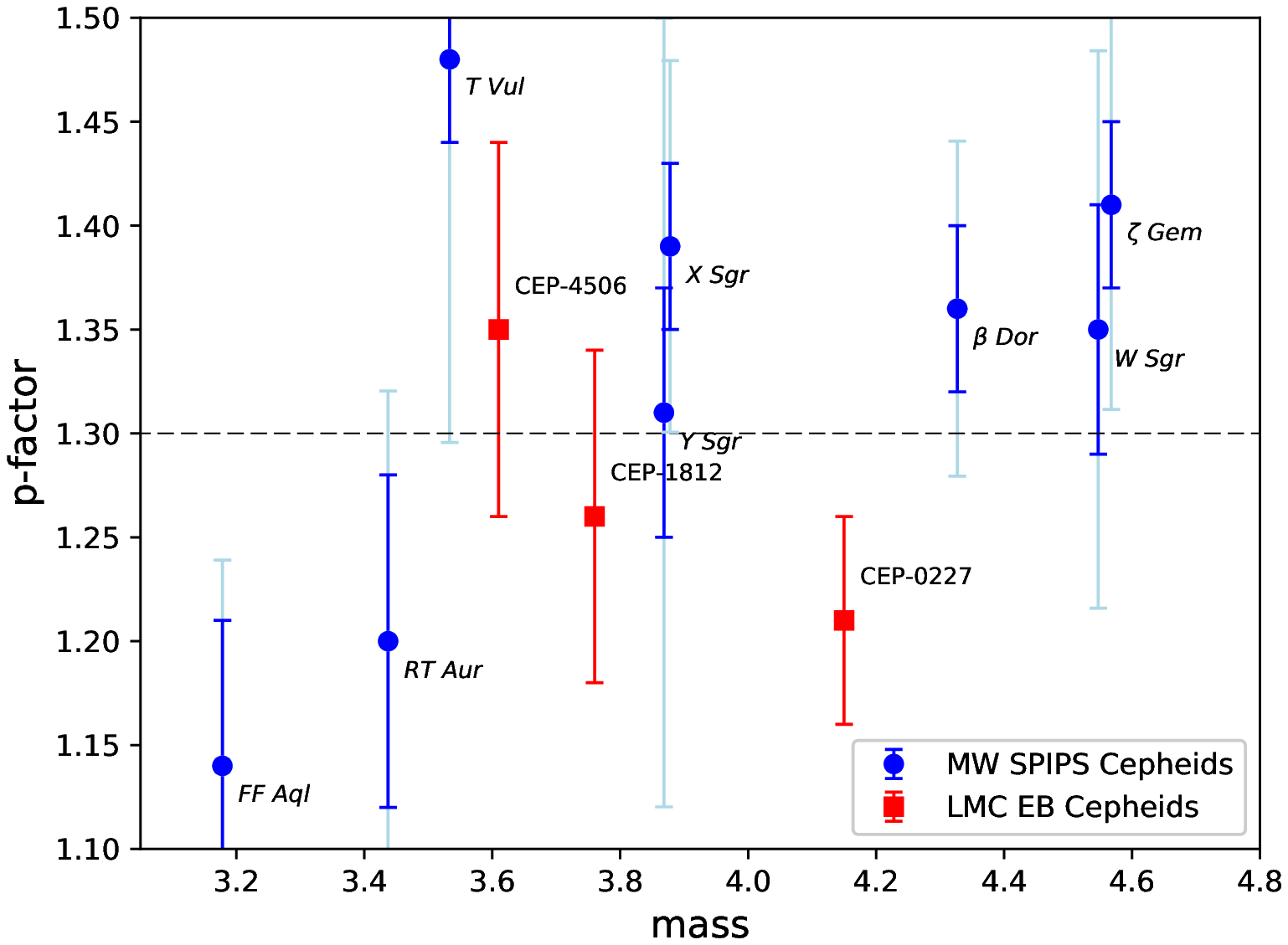}} 
  \resizebox{0.49\linewidth}{!}{\includegraphics{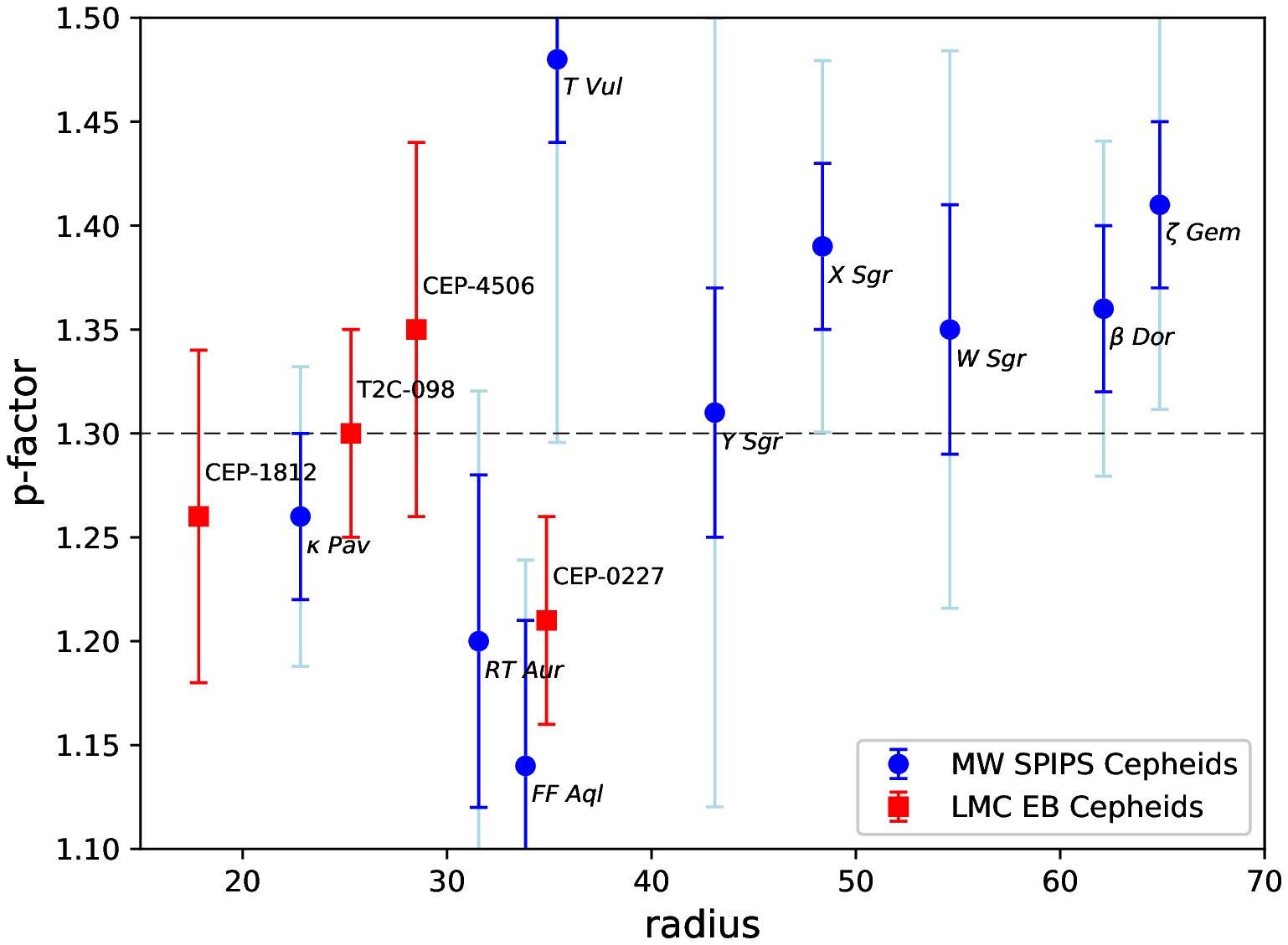}}
\caption{{\it left:} Mass -- p-factor diagram in the 3--5 $M_\odot$ range. {\it right:} Radius -- p-factor diagram. The uncertainties are marked as in Fig.~\ref{fig:pfac-per}. There is an interesting group of three stars with similar radii and low p-factors (RT Aur, FF Aql and CEP-0227). The high value of T Vul also looks exceptional, but for the moment the uncertainty is very high.
\label{fig:pfac-rel}}
\end{center}
\end{figure}

In Fig.~\ref{fig:pfac-per} one can still not see a clear period--p-factor relation, while a more structured and ordered distribution is seen in a radius--p-factor and mass--p-factor diagrams (in the latter low mass stars are also excluded), which are shown in Fig.~\ref{fig:pfac-rel}.  In the range of 3.5 to 4.2 $M_\odot$ indeed the results for Cepheids analyzed with the SPIPS code confirm the observations made above for our eclipsing binary Cepheids, but for the entire mass range displayed in the plot, i.e. 3 -- 5 $M_\odot$, the general trend is rather reversed -- higher mass Cepheids have higher p-factors. There is an interesting group of three stars with low p-factors and very similar radii -- RT Aur, FF Aql and LMC-CEP-0227 (see the right panel). The high p-factor value of T Vul, if confirmed, also looks exceptional.
Regarding the radius--p-factor diagram, one may also consider two separate groups, one with the p-factor increasing from LMC-CEP-1812 to T Vul and another starting from RT Aur and FF Aql and ending at $\zeta$ Gem. No significant correlation between the p-factor and $A_{RV}$ is seen for the Galactic Cepheids sample.

Although more studies and more data points, as well as higher-precision p-factors, will be necessary to obtain a definite answer, these results suggest that the p-factor dependence might not be so simple, and that additional parameters need to be included to correctly predict the p-factor value of a specific Cepheid. Perhaps the evolutionary state also needs to be considered.

\subsection{K-Terms for Cepheid Variables}

As the Cepheids in our study are members of binary systems we cannot study the K-term directly, but rather a differential K-term which depends on the parameters of both components. In our analysis, it is represented by the difference of $\gamma$-velocities of the components.
We have found that whenever the primary has higher $\log g$ than its companion, $\Delta\gamma$ is high, while when the companion has higher $\log g$, $\Delta\gamma$ is close to zero. This suggests that although the gravity may be the main factor, pulsations may also contribute to the effect, i.e. the K-term for the Cepheid is higher than the K-term for a stable star of similar gravity. This however may also be an effect of temperature, as the stable secondaries are in general cooler.

We have fitted a relation that depends on $\log g$ and T and obtained the following formula:
 $$ \Delta\gamma = 10 \times ( 1.13 \times (\log g_2 - \log g_1)  + 0.73\times(\log T_2 - \log T_1)). $$

We have also checked a relation based on only M and R, and obtained:
 $$ \Delta\gamma = 10 \times (-0.007 + 1.18 \times (\log M_2 - \log M_1)  - 0.94\times(\log R_2 - \log R_1)). $$

\begin{figure}
\begin{center}
  \resizebox{0.48\linewidth}{!}{\includegraphics{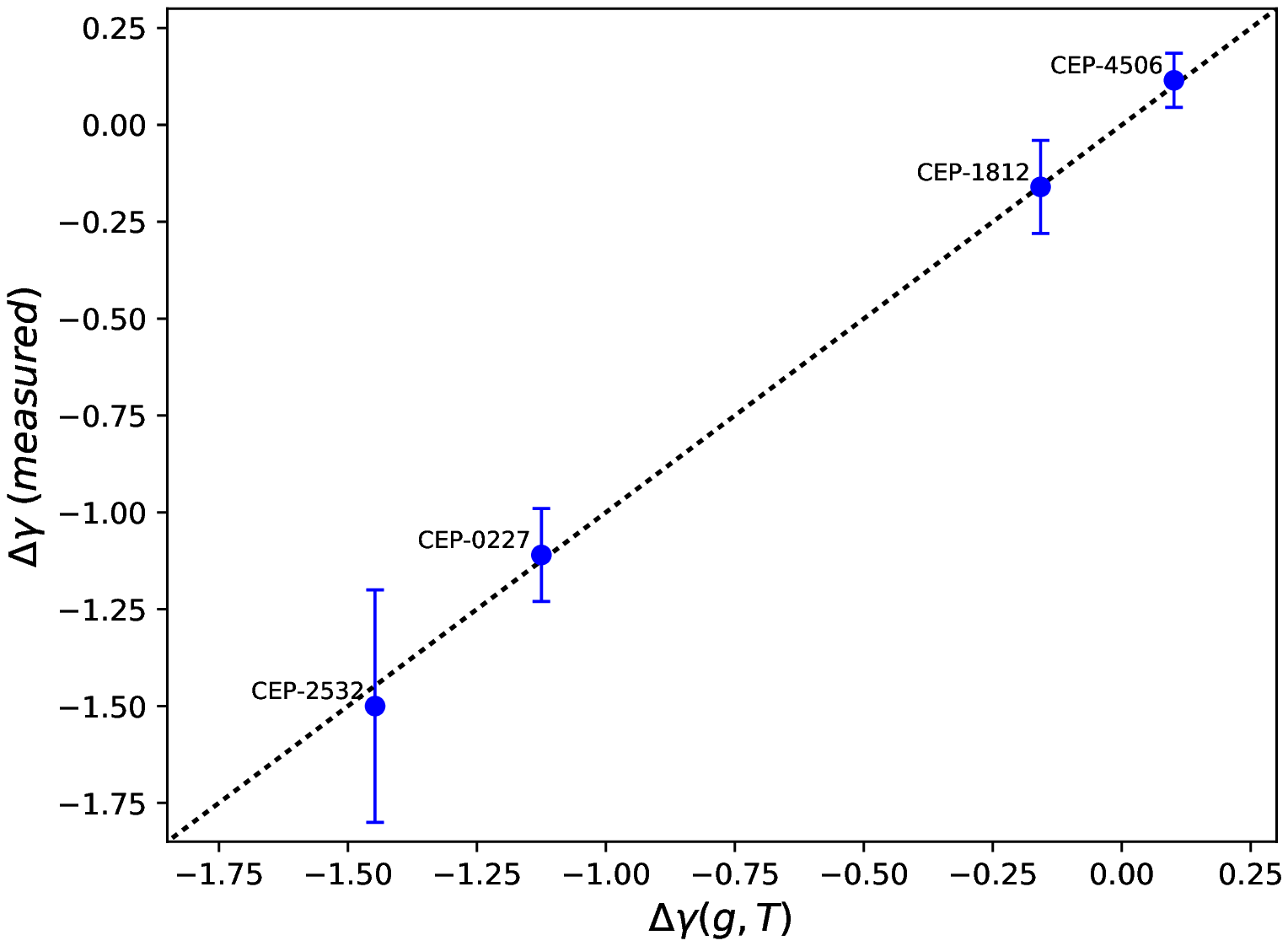}} 
  \resizebox{0.48\linewidth}{!}{\includegraphics{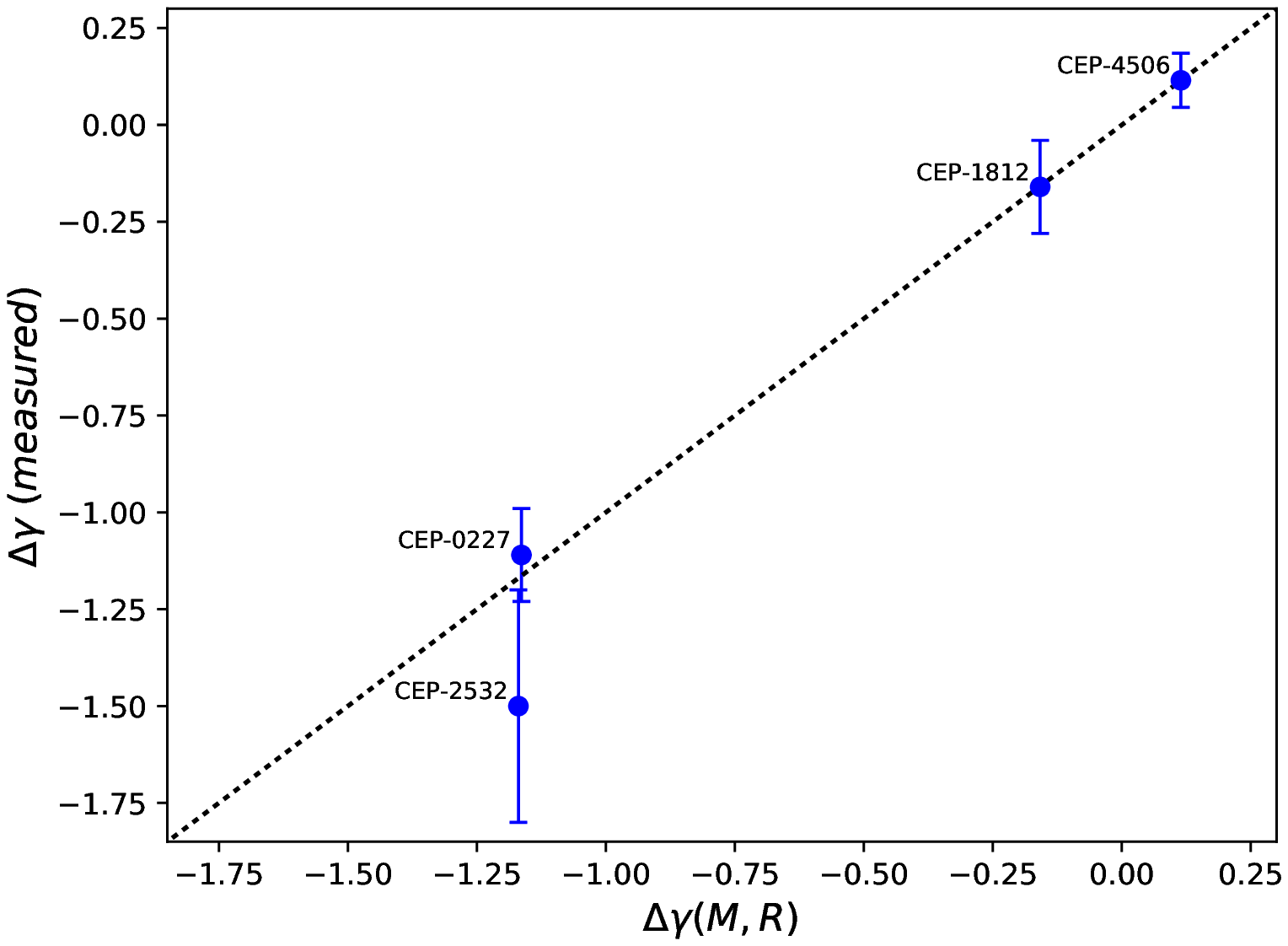}} 
\caption{K-term relation. Comparison of the difference between $\gamma$ velocities as measured and calculated from the physical parameters. {\it left:} $\Delta \gamma-g-T$ relation. {\it right:} $\Delta \gamma-M-R$ relation is also quite tight, but does not explain the very strong effect for LMC-CEP-2532, for which a temperature term is needed.
\label{fig:kterm-rel}}
\end{center}
\end{figure}

The relations are presented in Fig.~\ref{fig:kterm-rel}. Both work well, but the first  gives a better fit, as the latter does not explain well the very strong effect for LMC-CEP-2532, for which a temperature term is needed.

\section{Conclusions} \label{sec:conclusions}

Based on new spectroscopic and photometric data and updated models we have obtained improved physical parameters for seven previously analyzed pulsating stars, six of them classical Cepheids, all being components in eclipsing binary systems in the Large Magellanic Cloud.

The physical parameters of the Cepheids, particularly their masses and radii, are of a unique accuracy, typically of 1\%, and will lead to an improved understanding of stellar pulsation and the internal structure and stellar evolution of these stars. With this set of high quality mass, radius and temperature determinations for a sample of classical Cepheids at hand, we can start looking for relations that connect the parameters of the stars.
Below we present the main conclusions we have learned from the present analysis.

\begin{itemize}
\item All of our mass measurements for Cepheids are consistent with the expectations of pulsation theory.
\item Different solutions have been proposed for resolving the Cepheid mass discrepancy problem, and our precise measurements will contribute to understanding which parameters of non-canonical models lead to predictions of the evolutionary theory closest to the observations.
\item LMC-CEP-1718 is a particularly interesting system, the only one known to have two Cepheids orbiting each other. The masses of both Cepheids are now determined with an accuracy of 1\% each, and we  establish that the masses are slightly different, with the primary, shorter-period Cepheid being by about 1\% more massive than the longer-period Cepheid in the system, which has a larger radius. These characteristics make the evolutionary state of the components difficult to understand, unless some mass loss or enhanced evolution rate is taken into account.
\item At the moment the only anomaly we see for LMC-CEP-1812 is a radius which is slightly larger than expected, which however not justify any change in its classification yet. Its strange evolutionary state makes it however an interesting object for further study.
\item The evolutionary models of LMC-CEP-0227 should be updated, as all former studies assumed the companion to be the more massive and more advanced in its evolution. This is no longer valid as our new model strongly indicates that the Cepheid is slightly more massive than its companion.
LMC-CEP-0227 is a fundamental mode pulsator and its mass and radius are now determined with an accuracy of about 0.4\% each, making this classical Cepheid  the object with the most accurately known physical parameters of its class.
\item The new solution for LMC-CEP-4506  does not solve the problem of the non-pulsating secondary, which is  located well within the instability strip. More observations and a more sophisticated analysis may help to resolve or confirm this enigmatic finding.
\item From our mass and radius data, we have determined a precise period-mass-radius relation for classical Cepheids. Compared with theoretical period-mass-radius relations our result seems slightly more consistent with non-canonical models.
This empirical period-mass-radius relation can be used to estimate masses of Cepheids for which a dynamical mass cannot be directly measured (i.e. for single stars).
\item The mass of $\kappa$ Pav ($0.56 \pm 0.08 M_\odot$) derived using this relation may be treated as a first fully observational (though not direct) measurement of a mass of a type II Cepheid.
\item As expected, there is no difference between the first-overtone and fundamental mode classical Cepheids comparing radii and masses and their dependence on period. The only discriminant between the modes is temperature.
\item The p-factors measured for our sample show a significant scatter, indicating that some intrinsic dispersion is present and the results from the SPIPS method for a wider period range are consistent with our conclusions. Possible correlations with mass, radius and radial velocity amplitude are suggested, but the presence of more structured correlations also indicates that the p-factor may depend on many individual properties, and maybe even on the evolutionary state.
\item We found that to the first order the K-term is correlated with the gravity (or a combination of radius and mass in general) of the star, but that the pulsations and temperature may further enhance the effect.
\end{itemize}

Although a significant amount of data was already collected for all of these Cepheids, there are still many things we can learn from them. More advanced studies of the selected systems are planned once additional spectroscopic and photometric observations are  acquired.

\acknowledgments

We gratefully acknowledge financial support for this work from the Polish National Science Center grant MAESTRO 2012/06/A/ST9/00269 and from the BASAL Centro de Astrof{\'i}sica y Tecnolog{\'i}as Afines (CATA) PFB-06/2007. W.G. also acknowledges support for this work from the Chilean Ministry of Economy, Development and Tourism's Millenium Science Initiative through grant IC120009 awarded to the Millennium Institute of Astrophysics (MAS). B.P., M.T. and P.W. also acknowledge financial support for this work from the Polish National Science Center grant SONATA 2014/15/D/ST9/02248. The research leading to these results has received funding from the European Research Council (ERC) under the European Union's Horizon 2020 research and innovation program (grant agreement No 695099). The OGLE project has received funding from the National Science Centre, Poland, grant MAESTRO 2014/14/A/ST9/00121.

This work is based on observations collected at the European Organisation for Astronomical Research in the Southern Hemisphere under ESO programmes: 082.D-0499(A), 190.D-0237(C,E), 092.D-0363(A), 094.D-0056(A) and 097.D-0150(A). 
We would like to thank the support staff at the ESO Paranal and La Silla observatory and at the Las Campanas Observatory for their help in obtaining the observations, as well as for rest of the OGLE team for dedication of their time for collecting the photometric data. We thank ESO, Carnegie, and the CNTAC for generous allocation of observing time for this project.

The LMC background image used in Fig.\ref{fig:starsonsky} was created from the frames collected by the ASAS project led by Grzegorz Pojma\'{n}ski. B.P. would like to thank Richard I. Anderson for useful discussions and suggestions. We also thank the anonymous referee for the detailed comments that helped to improve this manuscript.

This research has made use of NASA's Astrophysics Data System Service.

\vspace{5mm}
\facilities{ESO:3.6m (HARPS), VLT:Kueyen (UVES), Magellan:Clay (MIKE)}

\software{
\texttt{JKTEBOP} \citep[][\url{http://www.astro.keele.ac.uk/~jkt/codes/jktebop.html}]{jktebop2004southworth},
\texttt{RaveSpan} \citep[][\url{https://users.camk.edu.pl/pilecki/ravespan/index.php}]{t2cep098apj2017},
}

\bibliographystyle{aasjournal}
\bibliography{cepheid_astrophysics}



\end{document}